\pdfoutput=1

\documentclass[%
 reprint,
amsmath,amssymb, aps,
superscriptaddress,
]{revtex4-2}

\usepackage{graphicx}
\usepackage{dcolumn}
\usepackage{bm}

\usepackage{color}

\begin{document}


\title{Active microphase separation in mixtures of microtubules and tip-accumulating molecular motors}

\author{Bezia Lemma}
\affiliation{Physics Department, Harvard University, Cambridge, MA 02138, USA}
\affiliation{Physics Department, Brandeis University, Waltham, MA 02453, USA}
\affiliation{Physics Department, University of California, Santa Barbara, CA 93106, USA}

\author{Noah P. Mitchell}
\affiliation{Kavli Institute for Theoretical Physics, University of California, Santa Barbara, CA 93106, USA}
\affiliation{Physics Department, University of California, Santa Barbara, CA 93106, USA}

\author{Radhika Subramanian}
\affiliation{Molecular Biology Department, Mass. General Hospital Boston, MA 02114, USA}
\affiliation{Genetics Department, Harvard Medical School, MA 02115, USA}

\author{Daniel J. Needleman}
\affiliation{John A. Paulson School of Engineering and Applied Sciences, Harvard University, Cambridge, MA 02138, USA}
\affiliation{Molecular \& Cellular Biology Department, Harvard University, Cambridge, MA 02138, USA}
\affiliation{Center for Computational Biology, Flatiron Institute, New York, NY 10010}

\author{Zvonimir Dogic}
\affiliation{Physics Department, University of California, Santa Barbara, CA 93106, USA}
\affiliation{Biomolecular Science \& Engineering Department, University of California, Santa Barbara, CA 93106, USA}
\affiliation{Physics Department, Brandeis University, Waltham, MA 02453, USA}
\email{zdogic@physics.ucsb.edu}
\date{\today}

\begin{abstract}
Mixtures of microtubules and molecular motors form active materials with diverse dynamical behaviors that vary based on their constituents' molecular properties. We map the non-equilibrium phase diagram of microtubules and tip-accumulating kinesin-4 molecular motors. We find that kinesin-4 can drive either global contractions or turbulent-like extensile dynamics, depending on the concentrations of both microtubules and a bundling agent. We also observe a range of spatially heterogeneous non-equilibrium phases, including finite-sized radial asters, 1D wormlike chains, extended 2D bilayers, and system-spanning 3D active foams. Finally, we describe intricate kinetic pathways that yield microphase separated structures and arise from the inherent frustration between the orientational order of filamentous microtubules and the positional order of tip-accumulating molecular motors. Our work shows that the form of active stresses and phases in cytoskeletal networks are not solely dictated by the properties of individual motors and filaments, but are also contingent on the constituent’s concentrations and spatial arrangement of motors on the filaments.
\end{abstract}

\maketitle


\section{\label{sec:intro}Introduction}
Active matter, the class of materials composed of motile energy-consuming units, exhibits various non-equilibrium dynamical phases \cite{marchetti2013rev, nedelec1997, schaller2010, bricard2013, narayan2007, soni2019}. For instance, active Brownian particles form dense clusters that share intriguing similarities with conventional gas-liquid phase coexistence, despite purely repulsive interactions \cite{Theurkauff2012, jeremie2013, aparna2013, cristina2012}. Active matter also exhibits distinct dynamical phases with no equilibrium analogs, such as percolating networks that undergo global contractions and turbulent-like flows observed in extensile cytoskeletal filaments or microscopic swimmers \cite{claire2018, dombrowski2004, tim2012, zhou2014, bendix2008, peter2015}. Theoretical tools that predict such macroscopic dynamics from microscopic details are still under development \cite{liverpool2005, gao2015, vliegenthart2020, belmonte2017, lenz2020}. Consequently, there is a lack of knowledge about the landscape of the possible dynamic phases that can arise in active matter systems. Our ability to rationally engineer large-scale dynamics by controlling the behavior of microscopic constituents is in its infancy \cite{ZDrev}. One way to address this critical knowledge gap is through experiments that measure detailed non-equilibrium phase diagrams of systems with varied microscopic dynamics.

Motivated by these considerations, we study the self-organization of microtubule filaments driven by tip-accumulating kinesin-4 molecular motors. We measure a non-equilibrium phase diagram, finding not only previously described contracting gels and extensile fluids, but also a range of novel structures, which include localized 1D micelle-like asters, extended 2D flat bilayers, monolayer covered condensates, and 3D bilayer-based foam-like networks. These structures are fundamentally different from previously studied forms of active matter, due to the importance of both positional and orientational order. Instead, they are more reminiscent of the diverse microphase-separated phases that self-assemble from chemically heterogeneous amphiphilic molecules~\cite{israelachvili1976, bates1990}. However, unlike equilibrium amphiphilic self-assembly, which is driven by the chemical immiscibility of different segments \cite{safran2018book}, the formation and continuous rearrangement of kinesin-4/microtubule structures are driven by energy-consuming molecular motors. We collectively name these phenomena \textit{active microphase separation}.

The dimeric kinesin-4 molecular motors used in this study consume energy from ATP hydrolysis to step towards microtubule plus ends, where they accumulate \cite{bieling2010, radhika2013, sithara2018}. Kinesin localization results in the formation of segmented microtubules consisting of a motor-rich segment at the microtubule plus end and an adjoining motor-poor segment. Thus, the unique properties of kinesin-4 motors yield a reconfigurable building block in which the motor dynamics encode the filament’s spatial heterogeneity, unlike the permanently encoded chemical structure of conventional amphiphiles. Microscopic parameters such as the microtubule length and the kinesin-4 concentration determine the size of the motor-rich domain. The plus-end segment can slide along other microtubules to their plus-ends \cite{sithara2018, sithara2020}. 

\section {Results}
\subsection{\label{sec:asters} Aster formation} 

\begin{figure*}
\includegraphics[width=\textwidth]{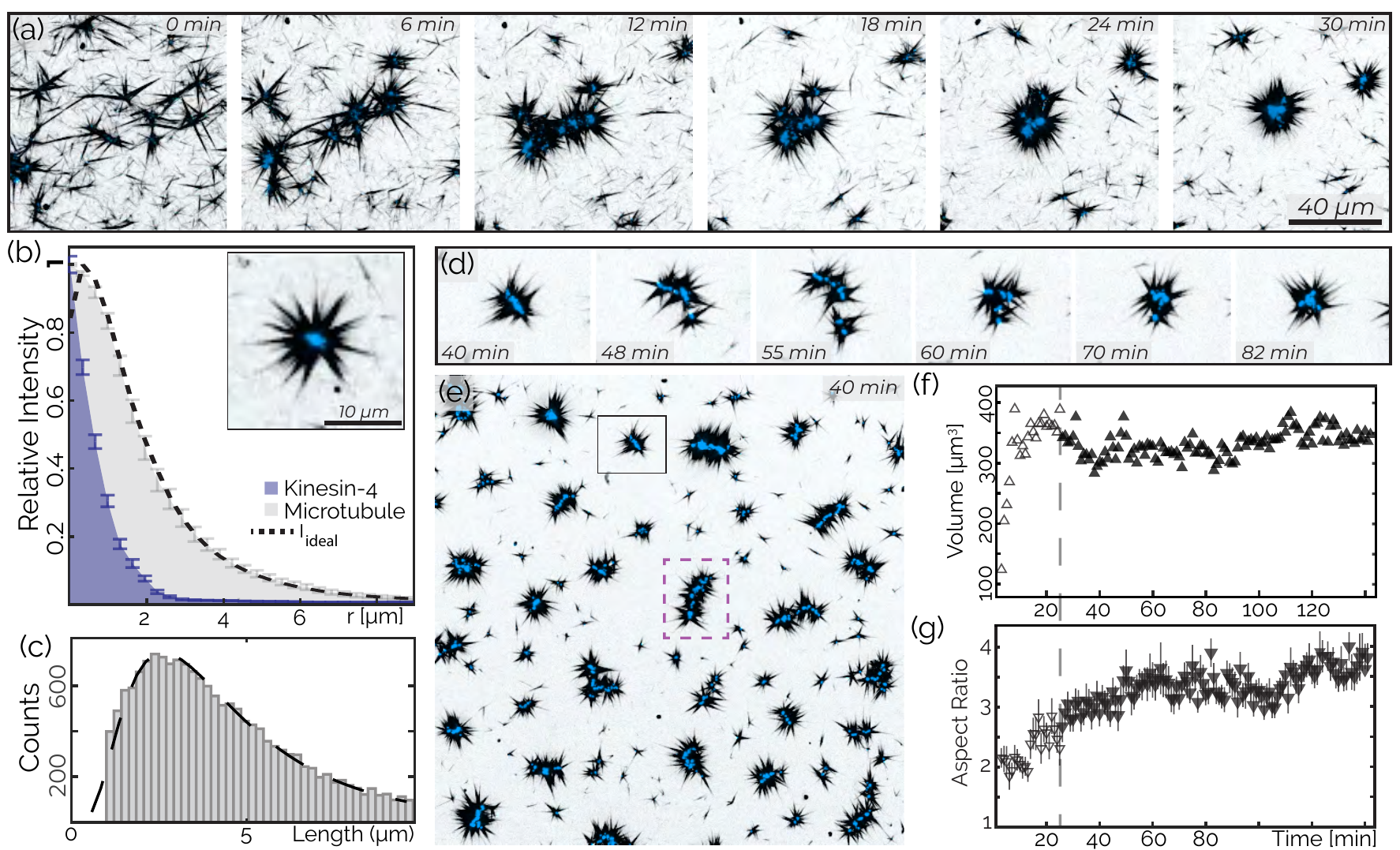}
\caption{\label{fig:aster} 
Self-organization of reconfiguring asters.
(a) Kinesin-4 induces rapid assembly of asters.
(b) The density profile of microtubules (gray) radially averaged from the z-projection of an aster. Predicted structures $I_{ideal}$ (dotted black line) based on end-bound kinesin-4 motors, given the measured density profile of kinesin-4 (blue). Bars are standard error averaged over three similar radial asters.
Inset: Aster with approximate radial symmetry. 
(c) Microtubule polydispersity (gray bars) is described by a log-normal distribution (dashed black line, M=1.4, S=0.6, mean 4.9 $\mu$m, mode 2.8 $\mu$m).
(d) Temporal rearrangement of an aster.
(e) A large field of view shows fully-formed asters. The dashed purple line highlights a wormlike structure.
(f) The mean aster volume as a function of time. Open shapes indicate the aster formation regime.
(g) The mean major/minor moment ratio of asters over time. Bars represent standard deviation.
All images are z-projections over 6.5 $\mu$m, sample contains 200 nM kinesin-4 (blue), 400 nM tubulin (black).}
\end{figure*}

We first studied the organization of a low concentration of stabilized microtubules by kinesin-4 motors in a thin parallelepiped chamber (See Methods). Immediately after mixing, we observed microtubules joined by their ends [Fig.~\ref{fig:aster}(a), 0 min]. Within the first $\sim$10 minutes, collections of microtubules continue to merge with each other, while labeled kinesin-4 clusters became visible at locations where filaments joined [Fig.~\ref{fig:aster}(a), 6-12 min]. Subsequently, the nascent kinesin clusters merged with each other, forming increasingly better-defined radial structures [Fig.~\ref{fig:aster}(a), 18-24 min]. The intensity of the motor-rich aster clusters located at the aster core increased, indicating a continual accumulation of motors. Within thirty minutes, the majority of microtubules condensed into radial star-shaped asters with well-defined kinesin-4 cores at their centers [Fig.~\ref{fig:aster}(a), 30 min].

To understand the aster structure, we measured the density profile of radially symmetric asters from 3D confocal images [Fig.~\ref{fig:aster}(b)]. The kinesin core had a radius of $\sim$1 $\mu$m, while the microtubule profile spanned $\sim$10 $\mu$m radially outwards. We hypothesized that microtubules were anchored to the aster core by their tips. To test this proposition, we modeled the aster’s structure by convolving the measured microtubule length distribution [Fig.~\ref{fig:aster}(c)] with the intensity profile of the kinesin core (SI). This convolution yielded a radially averaged microtubule profile that closely matched the experiments [Fig.~\ref{fig:aster}(b), dashed line], which is consistent with our hypothesis.

After their formation, asters continued to evolve by merging with each other and undergoing internal rearrangements [Fig.~\ref{fig:aster}(d)]. Over time this yielded elongated wormlike structures [Fig.~\ref{fig:aster}(e), Vid.~1]. To characterize such dynamics, we measured the mean three-dimensional moments of the kinesin-rich aster’s cores. The average ratio between the major and minor moments increased two-fold, while the mean volume of asters remained approximately constant [Fig.~\ref{fig:aster}(f), (g)].

\subsection{\label{sec:contraction} Global contraction and bilayer formation} 

\begin{figure*}
\includegraphics[width=\textwidth]{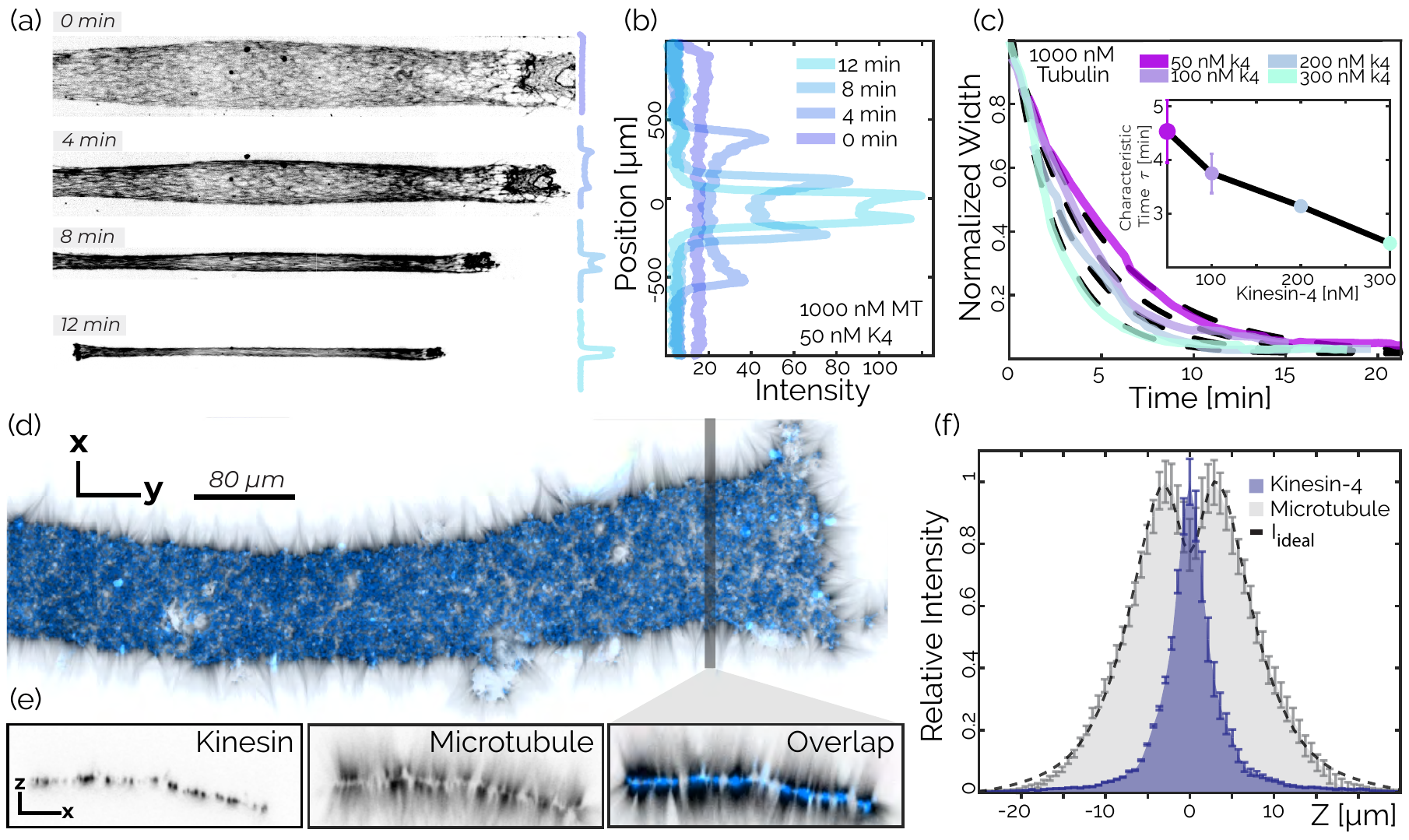}
\caption{\label{fig:globalcontract} Globally contracting networks generate bilayer structures.
(a) Kinesin-4 driven global contraction of labeled microtubules.
(b) Microtubule fluorescence as a function of position along the chamber’s short axis reveals non-uniform density growth, with peaks at the sample edges.
(c) The normalized width $W_n(t)$ of a contracting network decays over time. Dashed lines are fits of Eq.~\ref{eq:contract}.
Inset: Contraction timescale $\tau$ decreases with kinesin concentration. Error bars indicate standard error (n=3).
(d) The final structure of the contracted bilayer consists of a kinesin 2D sheet (blue) with microtubules (black) anchored to the surface and pointing along its normal.
(e) $x$-$z$ resliced at the shaded line.
(f) Fluorescence intensity profile along the surface normal. The predicted microtubule fluorescence $I_{ideal}$ (dotted black line) agrees with the measured fluorescence. Bars indicate standard error over twenty sections of 3 $\mu$m width.}
\end{figure*}

By increasing tubulin concentration above 1 $\mu$M, we observed the emergence of new dynamics. Instead of forming locally condensed asters, the system globally contracted into a single structure [Fig.~\ref{fig:globalcontract}(a), Vid. 2]. Material density was highest at the boundaries of the contracting network [Fig.~\ref{fig:globalcontract}(b)], similar to dynein-induced contractions studied in cell extracts and purified systems \cite{tan2018,peter2015}. We tracked the contracting network’s width $W(t)$ over time $t$. The normalized width, $W_n(t)=W(t)/W(0)$, was described by an exponential function:
\begin{equation} \label{eq:contract}
W_n(t) \approx W_n^\infty + e^{\frac{-(t-t_0)}{\tau}} (1 - W_n^\infty ) ,
\end{equation}
where $t_0$ is a time-offset, $W_n^\infty$ is the final normalized width, and $\tau$ is the contraction timescale [Fig.~\ref{fig:globalcontract}(c)]. $\tau$ increased with increasing kinesin concentration [Fig.~\ref{fig:globalcontract}(c)], and decreased with increasing microtubule number density [Fig.~\ref{fig:SuppContract}].

Examination of the final contracted state revealed a well-defined bilayer structure in which the kinesin motors formed an extended 2D sheet, with microtubules protruding from both sides of the sheet, pointing along the surface normal [Fig.~\ref{fig:globalcontract}(d), (e)]. In analogy to asters, we hypothesized that microtubules are anchored to the 2D kinesin sheet by their tips. We modeled the bilayer structure by convolving the measured length distribution of microtubules with the kinesin intensity profile along the surface normal (SI). The model of the bilayer structure closely matched the experimentally measured density profile [Fig.~\ref{fig:globalcontract}(f)]. Thus, our analysis suggests that microtubules are connected to the high-density kinesin layer by their plus-ends, with their minus-ends pointing outwards. How an initially disordered contracting network transforms into a late-stage bilayer structure remains to be studied. 

\begin{figure*}
\includegraphics[width=\textwidth]{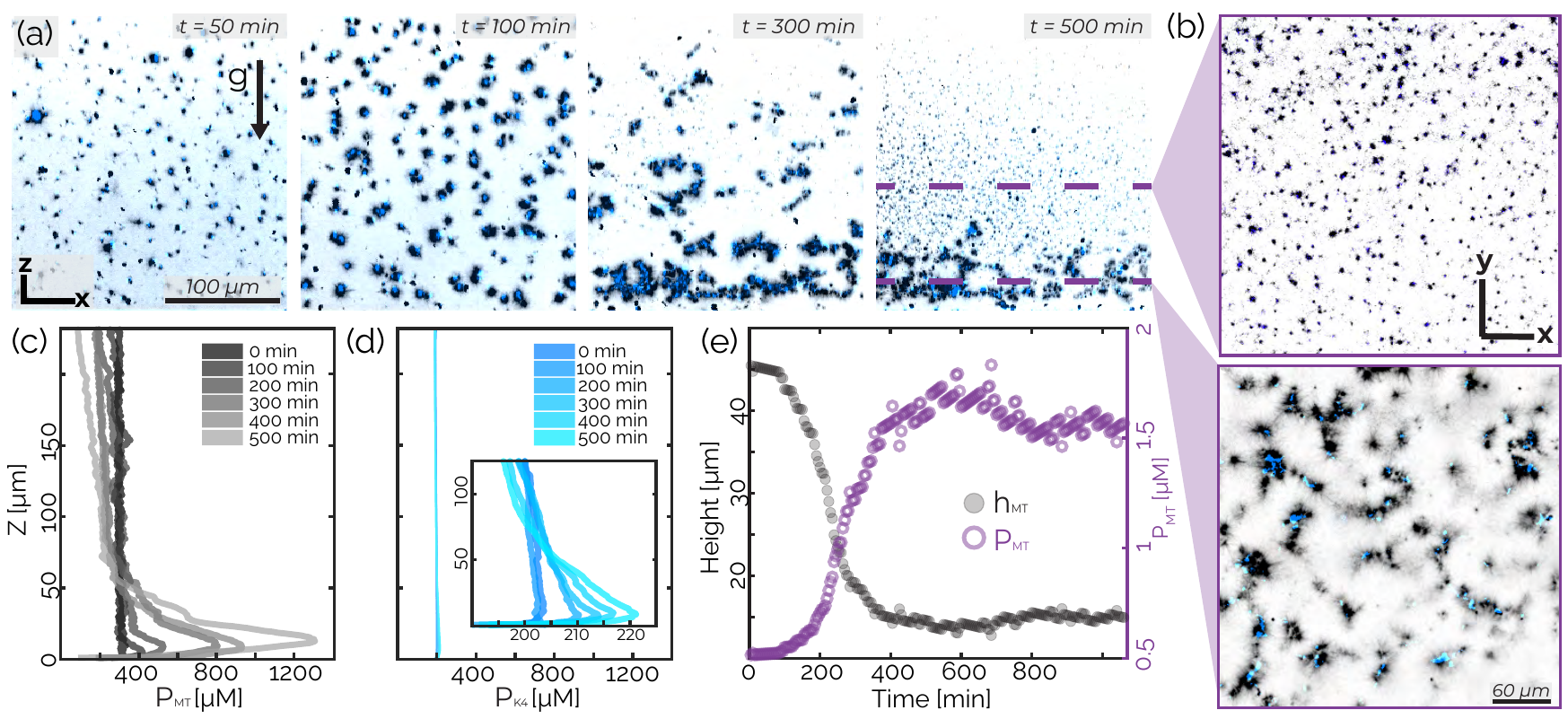}
\caption{\label{fig:sediment} Initial conditions determine steady-state dynamics.
(a) $x$-$z$ plane images show the aster assembly and sedimentation. The arrow indicates gravity, $x$-$y$ is the imaging plane.
(b) Asters images in the $x$-$y$ at two different heights at 500 min.
(c, d) Temporal evolution of the density $z$-profiles of microtubules $\rho_{MT}$ and kinesin $\rho_{K4}$ illustrate material sedimentation.
(e) The average microtubule density (purple open circles) below the sedimentation height (black circles) as a function of time. The effective tubulin concentration is higher than what is used in [Fig.~\ref{fig:globalcontract}] yet no global contraction occurs.}
\end{figure*}

We showed that increasing the microtubule concentration induces a transition from local asters to large-scale bilayers. To investigate the importance of initial conditions, we tested if increasing the concentration of fully formed asters leads to a similar transition. We prepared a sample with a low filament concentration in a tall sample chamber (250 $\mu$m), which led to the formation of asters throughout the volume. Once formed, large asters slowly sedimented into a dense $\sim$50 $\mu$m thick layer, which had an average tubulin density above 1 $\mu$M [Fig.~\ref{fig:sediment}(b-d)]. Uniformly dispersed samples prepared at such concentrations contracted into bilayers. However, the sedimented asters did not contract into a single structure. Instead, they formed a dense continuously rearranging network [Fig.~\ref{fig:sediment}(e), Vid. 3]. The lack of global contraction demonstrates that the form of the long-term steady-state structures depends not only on the constituents’ local concentration, but also on the sample history. Intriguingly, the increase in kinesin density due to sedimentation is an order of magnitude smaller than the increase in tubulin density [Fig.~\ref{fig:sediment}(d)]. Hence, in contrast to microtubules, a significant fraction of the kinesin does not incorporate into the asters. 

\subsection{\label{sec:roughening} Surface roughening of contracting networks}

\begin{figure*}
\includegraphics[width=\textwidth]{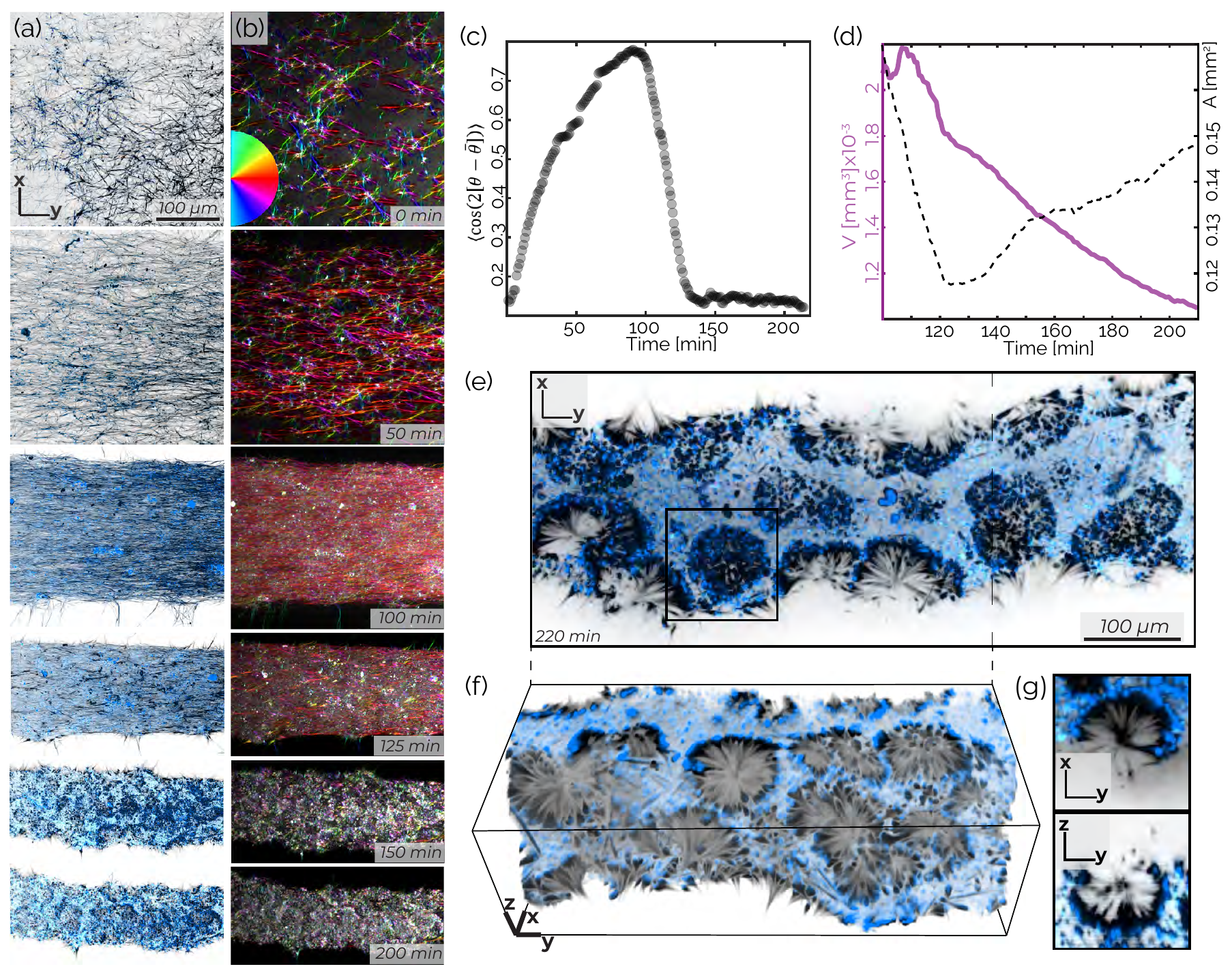}
\caption{\label{fig:highmtcontract} Nematic alignment and surface roughening of a contracting network.
(a) $z$-projected images demonstrate that decreasing network volume leads to increasing nematic alignment.
(b) $z$-projection of the microtubule nematic order. Hue indicates the nematic director indicated by the color wheel, while intensity indicates coherency (SI).
(c) The microtubule nematic order parameter increases during contraction and then decreases during roughening.
(d) The contracting network's volume (solid purple) decreases continuously. Its surface area (dashed black) initially decreases but then increases.
(e) A 10 $\mu$m z-projection of the material after surface roughening generates spherical cavities.
(f) A cropped 3D projection highlights the invaginated structure of the microtubule network.
(g) $x$-$y$ and $z$-$y$ show a hemispherical cavity.
Sample composed of 10 $\mu$M tubulin (black), 200 nM kinesin (blue). }
\end{figure*}

Samples prepared with even higher tubulin concentrations (10 $\mu$M) also underwent global contractions, but exhibited a distinct kinetic pathway and a different final structure from the above-described bilayers. The sample evolution proceeded in two stages: an initial global contraction followed by morphological surface roughening [Vid. 4]. In the first stage, the initially isotropic network developed nematic order while contracting [Fig.~\ref{fig:highmtcontract}(a)]. We defined $\theta$ as the local orientation of microtubule bundles in the structure's interior and $\bar{\theta}$ as the average bundle orientation [Fig.~\ref{fig:highmtcontract}(b), SI]. The scalar order parameter $S =\langle \cos(2[\theta -\bar{\theta}])\rangle$ indicates the degree of nematic ordering, with 0 representing isotropic structure and 1 representing perfect alignment (SI). As the network contracted, its volume $V$ decreased monotonically, while the order parameter $S$ of the enclosed microtubules increased [Fig.~\ref{fig:highmtcontract}(c)]. 

After approximately 120 minutes, the heretofore increasing nematic order parameter $S$ started decreasing sharply, signaling the onset of the second stage [Fig.~\ref{fig:highmtcontract}(c)]. Simultaneously, the network’s surface area $A$, which had previously fallen by a factor of two, began to increase [Fig.~\ref{fig:highmtcontract}(d)]. This transition was concomitant with morphological changes, in which the smooth interface of the contracting network started roughening. Surface roughening was accompanied by the formation of a dense monolayer consisting of a kinesin sheet with outwardly pointing microtubules, which enveloped the contracting network [Fig.~\ref{fig:highmtcontract}(e)]. Over time the roughening surface developed invaginations that rearranged into hemispherical $\sim$50 $\mu$m cavities [Fig.~\ref{fig:highmtcontract}(e), (f)]. Microtubules protruding from the surfaces of the hemispherical cavities reached the cavities' center, thus creating inverted asters with a sheet of kinesin half-enveloping radially splayed microtubules [Fig.~\ref{fig:highmtcontract}(g)]. 

\begin{figure*}
\includegraphics[width=\textwidth]{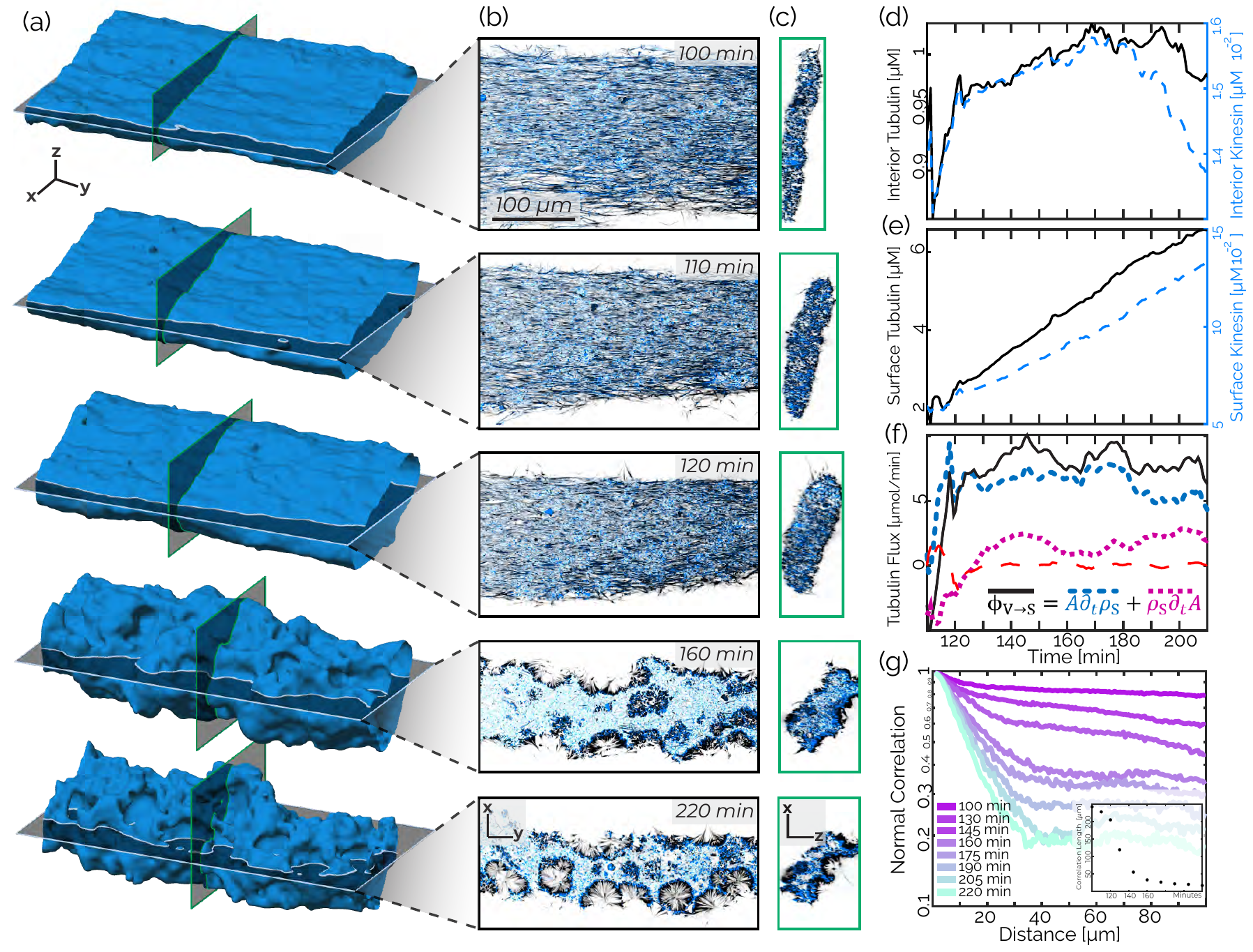}
\caption{\label{fig:3Dcontractflux} 
Surface roughening is accompanied by the formation of a surface-bound monolayer.
(a) Time series of a surface of a contracting network.
(b) $x$-$y$ slices of data corresponding to cuts shown in the previous panel reveal the formation of a monolayer and invaginations at late times.
(c) $x$-$z$ slices show contracting cross-section until the roughening commences.
(d) Tubulin and kinesin density within the interior of the contracting network is constant during the roughening phase.
(e) Tubulin and kinesin density within 5 $\mu$m of the surface increase during the roughening phase.
(f) The flux of microtubules from the interior to the surface $\Phi_{V \rightarrow S}$ (black solid), the microtubule surface density $A\partial_t \rho_s$ (blue dashed) and the change in surface area $\rho_s \partial_t A$ (purple short-dashed) as a function of time. The red long-dashed line indicates the sum of all three terms. 
(g) Normal-normal spatial correlations show faster decay as the material roughens. These correlations are calculated only on a bisected surface, to reduce the influence of the overall surface curvature.
Inset: Exponential fits to the normal-normal correlation decay between 10-20 $\mu$m show correlation length decreased by 200 $\mu$m over 50 minutes.
Sample consisted of 10 $\mu$M tubulin (black), 200 nM kinesin (blue). }
\end{figure*}

We reconstructed the network’s 3D structure using a morphological snakes level sets algorithm [Fig.~\ref{fig:3Dcontractflux}(a),(b)] \cite{marquez2014, chan2001, fedkiw2002book}. The surface and cross-sectional views show an initial rounding of the network’s cross-section, followed by a subsequent roughening [Fig.~\ref{fig:3Dcontractflux}(c)]. Numerical representation of the contracting network allowed us to quantify the distribution of the cytoskeletal material both on the surface and within the interior of the contracting network. During the second stage, while the density of the interior protein remained nearly constant [Fig.~\ref{fig:3Dcontractflux}(d)], the density of kinesin-4 and microtubules within 5 $\mu$m of the surface increased threefold [Fig.~\ref{fig:3Dcontractflux}(e)]. 

To understand whether the protein-dense shell arises simply from geometric deformation of the surface or by drawing material from the bulk, we quantified the kinematics of the partitioning between the dense network surface and its contracting interior.
In the roughening stage, the surface area $A$ increased [Fig.~\ref{fig:highmtcontract}(d)]. 
In the absence of any material flux between the surface and the interior, the areal density of surface-bound microtubules $\rho_S$ would decrease proportionally to the surface area growth: $A \partial_t \langle \rho_S \rangle = - \langle \rho_S \rangle \partial_t A $ (SI). We find that these two terms are, in fact, far from equal and opposite [Fig.~\ref{fig:3Dcontractflux}(f)], suggesting that there is substantial flux from the interior to the surface. 
Meanwhile, the sum total of all microtubule fluorescence is constant. 
The implied mass conservation is described by
\begin{equation}
A \partial_t \langle \rho_S \rangle + \langle \rho_S \rangle \partial_t A = \Phi_{V\rightarrow S}, 
\end{equation}
where $\Phi_{V\rightarrow S}$ is flux of material from the interior to the surface. 
We then independently measured the flux of microtubules leaving the interior of the contracting network,
\begin{equation}
\Phi_{V\rightarrow S}= -V\partial_t \langle \rho_V \rangle - \langle \rho_V \rangle \partial_t V,
\end{equation}
where $\langle\rho_V\rangle$ is the average volumetric density of microtubules and $V$ is the volume of the interior, and find that it quantitatively accounts for the increasing density of the surface-bound microtubules $A \partial_t \langle \rho_S \rangle$ [Fig.~\ref{fig:3Dcontractflux}(f)]. Our analysis reveals that the density change due to surface area increase $\langle \rho_S \rangle \partial_t A$ is small compared to the mass transfer due to the flux from the interior to the surface $\Phi_{V \rightarrow S}$. The mechanism that drives the flux of microtubule transport from the interior to the surface remains unknown. 

To quantify the roughening transition, we measured the spatial correlations of the surface normals. A normal vector $\hat{n}(r,t)$ describes the network at each surface point $r$ at time $t$ (SI). The averaged correlation between all normal vectors, separated by a geodesic of length $\Lambda$, is given by
\begin{equation}
C(\Lambda,t) = \frac{\langle \hat{n} (r,t) \cdot \hat{n} (r+\Lambda,t)\rangle }{\langle \hat{n}(r,t)\cdot \hat{n}(r,t)\rangle } ,
\end{equation}
where angular brackets indicate a spatial average over all initial points and all geodesic paths of length $\Lambda$. At the beginning of the roughening stage, the network has an extended flat shape which reflects the chamber geometry. When restricted to either the top or bottom of the surface, pairs of normal vectors $\hat{n}$ point in similar directions even at large distances. Consequently, $C(\Lambda,t)$ remains close to unity for all values of $\Lambda$. As the surface roughens with time, the correlation between surface normals $\hat{n}$ decreases. $C(\Lambda,t)$ develops a plateau at large distances, where the plateau magnitude decreases with time [Fig.~\ref{fig:3Dcontractflux}(g)]. At smaller length scales, ranging from 1 to 30 $\mu$m, $C(\Lambda,t)$ exhibits exponential decay. The rate of the exponential increased six-fold from the beginning to the end of the roughening process. The long-range normal-normal correlation decayed from $C$($40$ $\mu$m, $100$ min) $\approx0.85$ to $C$($40$ $\mu$m, $220$ min) $\approx0.2$. 

\subsection{\label{sec:foam} Active foam formation}
\begin{figure*}
\includegraphics[width=\textwidth]{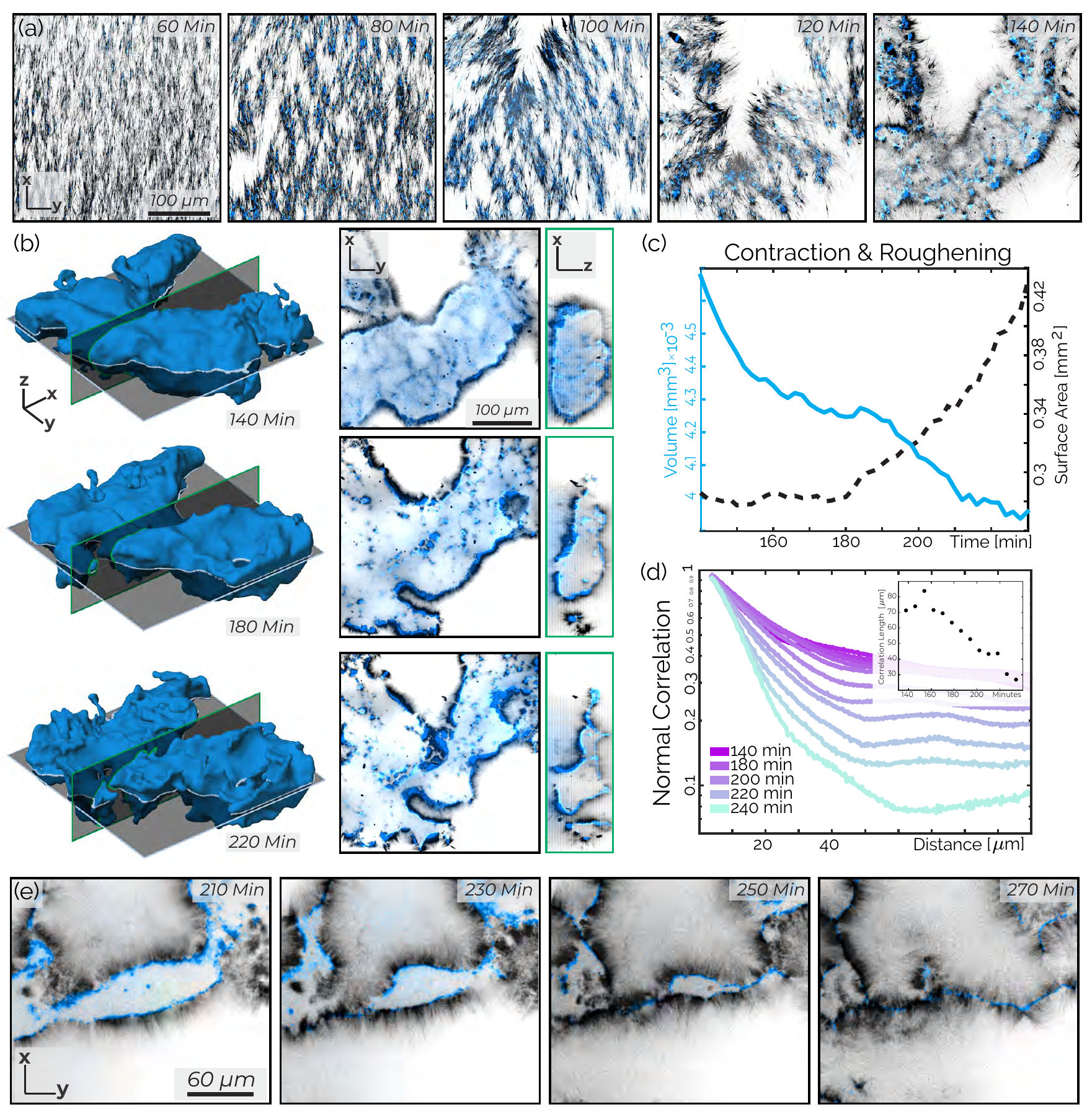}
\caption{\label{fig:3Dcoral} 
Splay-like deformations, self-tearing, and roughening at the highest microtubule concentrations.
(a) Maximum intensity $z$-projections over 3 $\mu$m show a splay-like instability that generates density variation and self-tearing that yields condensates.
(b) Evolution of a contracting condensate surface (left) $x$-$y$ and $x$-$z$ image cross-sections (right).
(c) The volume (solid blue curve) and surface area (black dashed curve) of a contracting condensate as a function of time.
(d) The spatial correlation between surface normal vectors decay over time.
Inset: Exponential fits to the normal-normal correlation decay between 5-20 $\mu$m show correlation length decreased by 50 $\mu$m over 80 minutes.
(e) Two surface-bound monolayers zippering into a bilayer.
Sample contained 200 nM kinesin (blue), 40 $\mu$M tubulin (black). }
\end{figure*}

At the highest tubulin concentrations studied (40 $\mu$M) we observed a multistage kinetic pathway of significant complexity [Vid. 5]. In this regime, the microtubules had an initial orientational order, and initially displayed subtle bend deformations [Fig.~\ref{fig:SuppFoamEvolution}]. Subsequently, the buckling dynamics transitioned into more dramatic splay-like deformations, the onset of which broke up the continuous network by generating sharp density variations between filament-rich and filament-poor regions [Fig.~\ref{fig:3Dcoral}(a), 80 min]. These changes in orientational order and local density fluctuations yielded finite-sized condensates that were well-separated from a background fluid mostly devoid of protein [Fig.~\ref{fig:3Dcoral}(a), 140 min]. A high-density monolayer of kinesin and microtubules enveloped the condensate surface, with microtubules aligned along the surface normal. The monolayer-covered condensates were similar to those observed at lower filament concentrations. The main difference is that active stresses ruptured the network, creating finite-sized structures. In contrast, lower microtubule concentrations generated only one contracting network, which did not break apart.

After their formation, condensates exhibited surface roughening. Using the previously described algorithm, we numerically generated surfaces describing the evolution of the condensate’s morphology [Fig.~\ref{fig:3Dcoral}(b)]. The condensate's volume decreased continuously, while its surface area $A$ remained constant until $\sim$160 minutes, after which $A$ increased sharply [Fig.~\ref{fig:3Dcoral}(c)]. As roughening continued, the mean curvature increased, and the normal-normal correlation $C(r)$ decreased [Fig.~\ref{fig:3Dcoral}(d), Fig.~\ref{fig:SuppMean}]. High-resolution images revealed the macroscopic mechanism driving the roughening transition. Crumpling monolayers encountered each other, generating a zippering transition of the kinesin decorated surfaces which locally produced a well-defined bilayer [Fig.~\ref{fig:3Dcoral}(e)].

\begin{figure*}
\includegraphics[width=\textwidth]{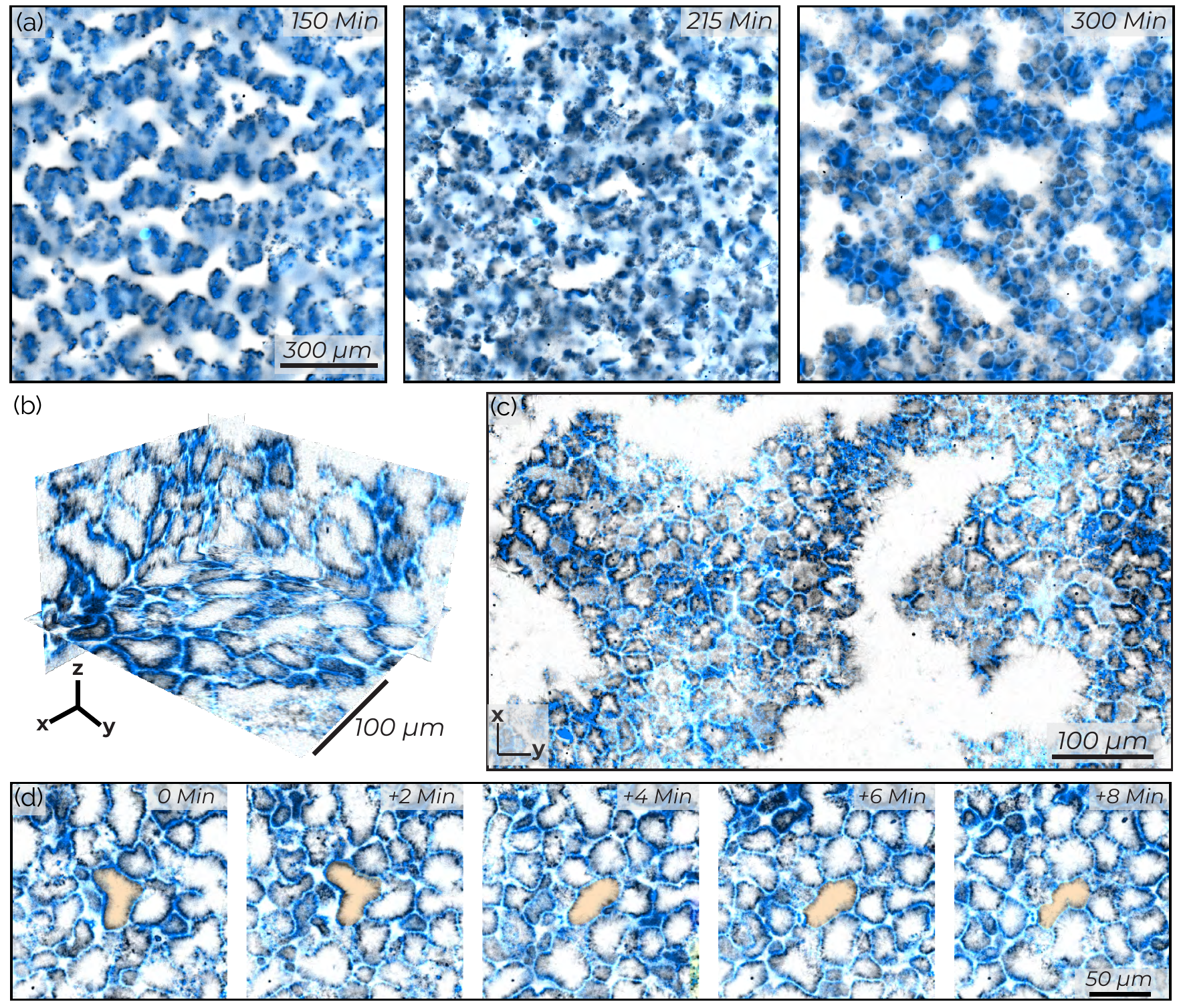}
\caption{\label{fig:timecoral} Surface roughening yields an active foam.
(a) Morphological change from monolayer envelopes to a percolated foam.
(b) Ortho-slices show the complex 3D structure of the active foam.
(c) Maximum intensity $z$-projection over 10 $\mu$m illustrates distinct foam cells which can have free ends or open faces.
(d) A foam cell undergoes topological rearrangements in an active foam.
Samples constituted from 200 nM kinesin (blue), 40 $\mu$M tubulin (black).}
\end{figure*}

On long times, the surface roughening transition generated an active foam, which consists of a 3D network of bilayers that connect through junctions. [Fig.~\ref{fig:timecoral}(a), Fig.~\ref{fig:SuppFoamEvolution}]. As in conventional foam, the interconnected bilayer surfaces formed cells, which had elongated or even winding shapes [Fig.~\ref{fig:timecoral}(b), Fig.~\ref{fig:SuppBigFoam}]. Unlike conventional foams, cells in an active foam had open sides, while the constituent bilayers had free-standing edges [Fig.~\ref{fig:timecoral}(c), Fig.~\ref{fig:SuppBigFoam}(b)]. The borders of the active foam compartments consist of microtubule/kinesin-4 bilayers [Fig.~\ref{fig:timecoral}(b), (c)]. The active foam exhibited topological rearrangements. Individual cells deformed, while bilayer walls moved to change the local topology [Fig.~\ref{fig:timecoral}(d), Vid. 6]. Thus, the surface roughening transition is the first stage of a unique morphological transition in which a continuous and smooth space-filling condensate transforms into perforated foam-like structures. The development of an active foam and its rearrangements remains an important topic for future investigations.

\subsection{\label{sec:gels} A bundling-induced transition from contracting to extensile gels}
\begin{figure*}
\includegraphics[width=\textwidth]{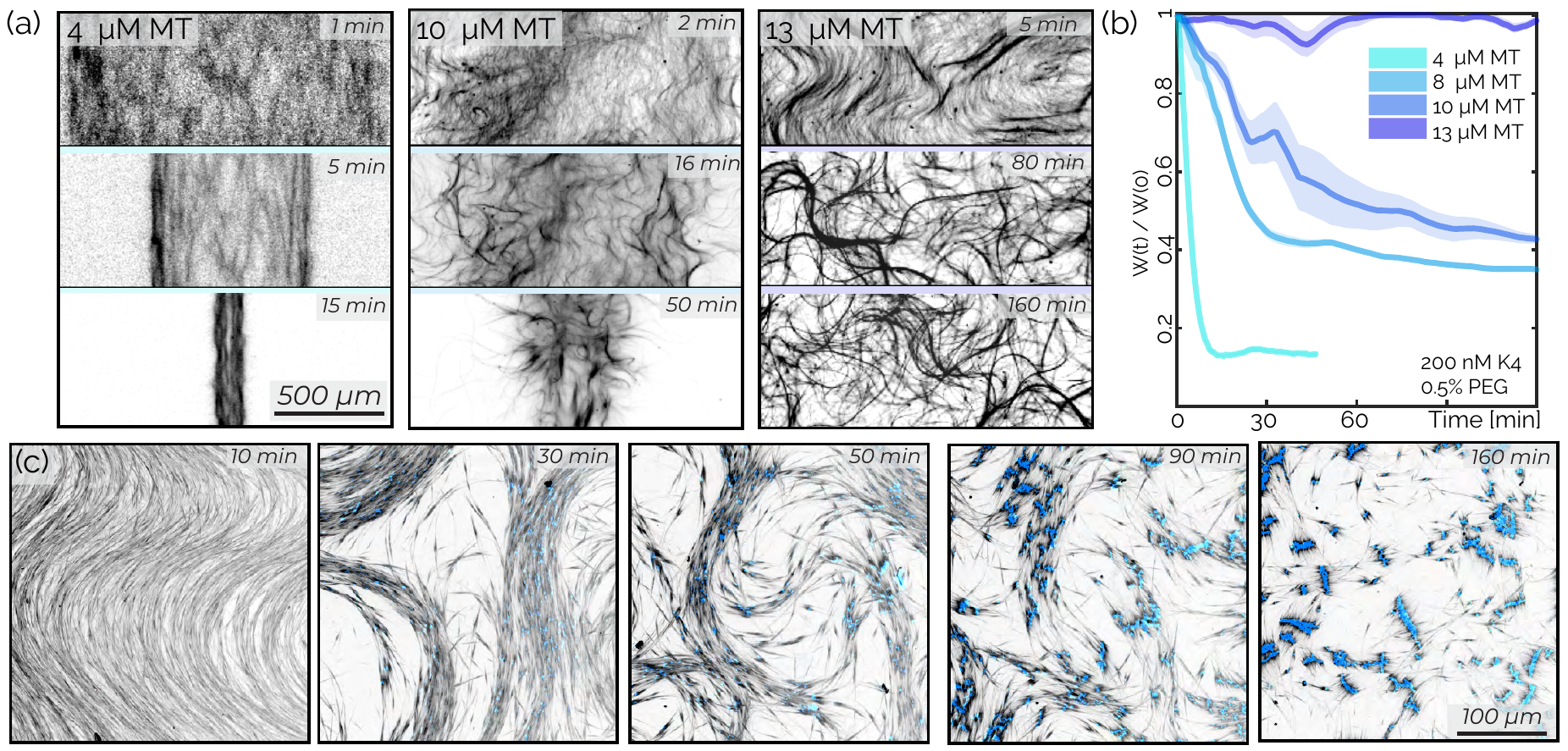}
\caption{\label{fig:gels} Microtubule bundling yields extensile dynamics. 
(a) The evolution of the shear-aligned microtubule network depends on filament concentrations. Samples had 0.5\% PEG, 300 nM kinesin. 
(b) The average microtubule network width $W(t)$, normalized by the initial width $W(0)$, decreased over time, with lower microtubule densities contracting faster. The shaded region indicates the standard deviation from data taken at five non-overlapping positions over the long axis of the chamber. 
(c) Extensile instability leads to the formation of a bilayer structure. This sample chamber was 30 $\mu$m thick this sample contained 100 nM kinesin (blue), 80 $\mu$M tubulin (black) and 0.1\% PEG.}
\end{figure*}

In the work described so far, we observed local and global contractions with increasing microtubule concentrations. In comparison, kinesin-1 generates extensile stresses when microtubules are combined with a microtubule bundling agent \cite{tim2012,pooja2018}. To investigate the capability of kinesin-4 motors to generate extensile stresses, we added a non-adsorbing polymer, PEG (polyethylene glycol), which bundles microtubules while still allowing for their relative motor-driven sliding \cite{andy2015}. At low microtubule concentrations (4 $\mu$M), global contractions occurred even in the presence of 0.5\% w/w PEG [Fig.~\ref{fig:gels}(a)]. However, beyond a critical filament concentration (10 $\mu$M tubulin), the material exhibited initial self-generated bend-like patterns which are suggestive of extensile stresses [Vid. 7] \cite{marchetti2013rev,ramaswamy2010}. On longer times scales, these materials did not contract but rather yielded a continuously rearranging network, similar to those previously studied [Fig.~\ref{fig:gels}(a)]~\cite{pooja2020,gil2014}.
The contractile to extensile transition was quantified by plotting the final network width $W(t)$ [Fig.~\ref{fig:gels}(b)]. At low filament concentrations, $W(t)$ monotonically decreases and then plateaus, characteristic of contraction. Increasing microtubule concentration further resulted in a network that spanned the entire chamber while continuously rearranging. Therefore $W(t)$, did not change over time. Using particle image velocimetry, we found that the mean microtubule network speed increased with increasing kinesin concentration. In contrast to kinesin-1 studies, increasing kinesin-4 concentration increased the velocity-velocity correlation length scale [SI] \cite{gil2014}.

We also observed that extensile gels could transform into globally contracted bilayers [Fig.~\ref{fig:gels}(c), Vid. 8]. Upon preparation, an active mixture (0.1-0.3\% w/w PEG, 80-90 $\mu$M tubulin) exhibited a bend instability and fluidized. However, on longer time scales, distinct segments of kinesin-4 appeared. As these segments became prominent, the motor driven dynamics slowed down. This dynamical transition was concomitant with the appearance of local bilayer-like arrangements. In these bilayers, kinesin-4 formed a central line with microtubules pointing outward on both sides.

\subsection{\label{sec:phases} A non-equilibrium phase diagram}
\begin{figure*}
\includegraphics[width=\textwidth]{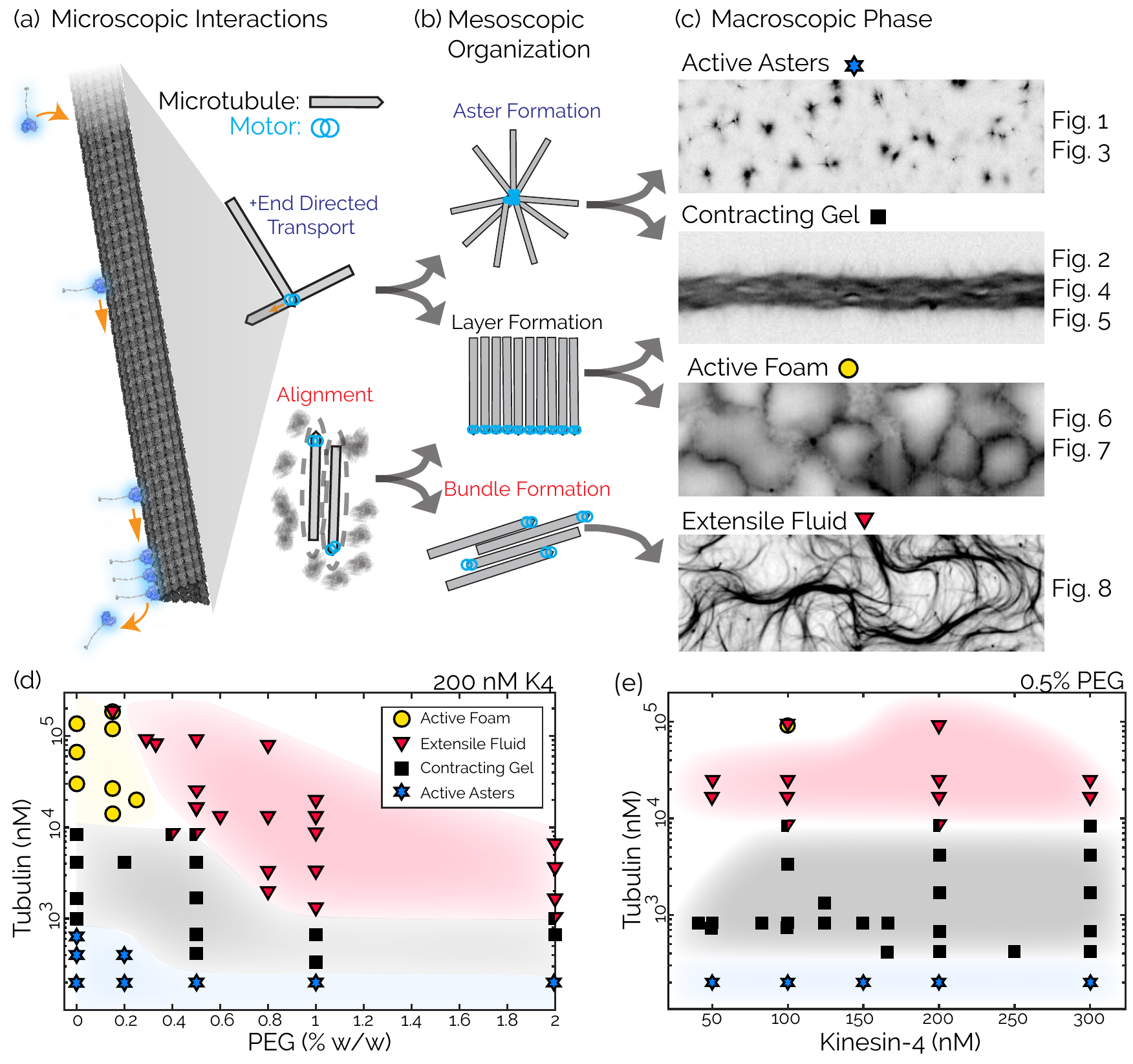}
\caption{\label{fig:phase}
The phase diagram of kinesin-4 and microtubules.
(a) Microscopic building blocks: kinesin-4 (blue) attach to a microtubule (grey), walk to the microtubules plus end, and accumulate at the plus end, creating a heterogeneous filament that can interact with other filaments by directed transport or via steric alignment induced by PEG.
(b) Mesoscale organizational motifs include asters, layers, or bundles.
(c) Hierarchically organized mesoscale building blocks yield macroscopic phases including dynamic asters, globally contracting gels, active bilayer foams, and fluidized extensile bundles.
(d) Phase diagram at 200 nM kinesin as a function of tubulin and PEG concentration.
(e) Phase diagram at 0.5\% PEG (w/w) as a function of protein concentrations.}
\end{figure*}

As described above, a one-dimensional sweep of tubulin concentration in the absence of PEG yielded active microphase separated phases, while adding PEG produced an active extensile fluid. To further characterize the system, we mapped the non-equilibrium phase diagram by creating samples between 50 and 300 nM kinesin-4, 0.2 to 180 $\mu$M tubulin, and 0\% to 2\% PEG [Fig.~\ref{fig:phase}(d),(e)]. At relatively low microtubule concentrations, the active material contracted into localized asters over a wide range of PEG and kinesin-4 concentrations. Increasing microtubule concentration generated global contractions, again over a wide range of PEG and kinesin-4 concentrations. At the highest microtubule concentrations, with little or no PEG, we observed the formation of active foams. Adding PEG in this regime transformed active foams into extensile turbulent-like gels similar to those seen in kinesin-1 driven systems. Presumably, introducing PEG suppressed the formation of asters and bilayer foams, while promoting the formation of bundles that generate extensile dynamics [Fig.~\ref{fig:phase}(d)]. Kinesin-4 concentration determined the speed of the autonomous dynamics but did not substantially affect the boundaries between the extensile and contracting phases [Fig.~\ref{fig:phase}(e), SI]. The long-term non-equilibrium phase behavior described here depends on the initial and boundary conditions, the sample history, and the kinetic pathways [Fig.~\ref{fig:SuppBoundaryConditions}].

\section{\label{sec:discussion} Discussion}
In cytoskeletal active matter, extensile active stresses drive continuous turbulent-like flows, while isotropic contracting active stresses generate local or global collapse \cite{ZDrev, tim2012, peter2015, murrell2012, tan2018, e2011activ, stam2017, kumar2018}. We studied the self-organization of microtubules and kinesin-4, a tip-accumulating molecular motor. In the regime of high concentrations of filaments and bundling agents, we observed extensile turbulent flows. Reducing either the concentrations of microtubules or PEG resulted in contraction. These observations demonstrate that the form of the active stress is not solely dictated by the molecular properties of cytoskeletal components, but is also dependent on the concentration of the constituents. This insight is valuable for relating the mesoscopic active stresses to the structure, interactions, and dynamics of the microscopic constituents \cite{zhang2021, blackwell2016, belmonte2017, ronceray2016}. In the contracting regime, we observed a myriad of active microphase separated structures. Lowest filament concentration sample yielded isolated asters [Fig.~\ref{fig:aster}]. With increasing filament concentrations, asters transformed into 1D wormlike structures, extended 2D bilayers, and foam-like 3D material [Fig.~\ref{fig:globalcontract},~\ref{fig:timecoral}]. These findings have implications for our understanding of cytoskeletal active matter.

The formation of aster-like structures has previously been observed in mixtures of microtubules and various molecular motors \cite{nedelec1997, hentrich2010, surrey2001, kruse2004, peter2015, husain2017}. Theoretical models of such asters are sometimes couched in the language of topological defects in liquid crystals. However, the asters studied here are well-isolated structures in a filament-free background fluid; thus they are more reminiscent of equilibrium amphiphile-based micelles. Instead of hydrophobic interactions, their condensation is driven by tip-accumulating molecular motors. With increasing concentration, amphiphilic systems form 1D wormlike micelles, 2D membranes and space-filling 3D lamellar, hexagonal, or disordered gyroid phases \cite{safran2018book}. We observed active analogs of these higher-order phases. Once the microphase separation is complete, motors continue to reconfigure the material, as we observed for both wormlike structures and active foams [Vid. 1,3,6]. Kinesin-4 drives these large-scale events by generating active stresses that are likely distinct from those postulated for a suspension of aligned active filaments. 

Molecular motors can mediate different filament interactions. For example, they can drive interfilament sliding within an aligned bundle, or they can cluster tips of isotropically arranged filaments \cite{sithara2018, peter2015, palenzuela2020}. Clusters of kinesin-1 motors are thought to primarily induce filament sliding \cite{gil2014}. However, observation of asters in such systems suggests that they retain a small degree of end-binding \cite{surrey2001}. In comparison, kinesin-4 has an enhanced end-binding property, which has been characterized on the single filament level \cite{sithara2018, sithara2020}. We developed a model of aster structure that predicts the microtubule profile from a given kinesin profile, but it does not explain the size of the kinesin core. The latter could be related to the size of the kinesin-4 cap. More experimentation is needed to elucidate this point, as single filament experiments suggest that the cap size depends on protein concentrations and microtubule length \cite{sithara2020}. Thus, the balance of spatial filament decoration and interfilament sliding by molecular motors might determine the range of possible phases of an active cytoskeletal material, and is a promising avenue for further investigation.

Active microphase separation has relevance to biological systems. The self-organization of microtubules and molecular motors have been studied in Xenopus egg extracts \cite{hannak2006, thawani2019}. Dynein drives aster assembly in Xenopus egg extracts, which globally contract at higher filament concentrations \cite{pelletier2020, tan2018, verde1991}. Such asters have been used as models for spindle pole assembly \cite{verde1991}. Under other conditions, stabilized microtubules in Xenopus egg extracts assemble into structures reminiscent of the bilayers observed in the present work \cite{mitchison2013}. In these experiments, extended bilayers of taxol stabilized microtubules form, with their minus ends pointing away from the midplane. These bilayer structures serve as models for the spindle midzone, the array of microtubules that assembles between segregating chromosomes and drives the spindle elongation and chromosome separation \cite{scholey2016, maya2021rev, chehang2019}. Much prior work on spindle midzones focused on factors that determine the extent of antiparallel overlap of the microtubule ends \cite{sithara2018, hannabuss2019}. However, the reason why this narrow region of antiparallel overlap stays well aligned across the entire spindle width remains poorly understood. The similarity between the bilayers observed in the present work, those formed in Xenopus egg extracts, and the spindle midzone itself, suggests that similar principles might govern the self-organization of all of these structures.

Besides revealing a range of active microphase states, our work also demonstrates rich kinetic pathways that lead to the formation of these phases. These pathways are influenced by the interplay between the tendency of rod-like filaments to align due to excluded volume interactions and the propensity of tip-adhering kinesin motors to drive microphase separation. We observe filament alignment at high microtubule concentrations, which occurs either initially during sample loading, or develops over time in a contracting network [Fig.~\ref{fig:highmtcontract},~\ref{fig:SuppFoamEvolution}]. Theory dictates that aligned active filaments are inherently unstable \cite{simha2002}. Specifically, extensile active stresses drive the bend instability as we observed for the kinesin-4 system in the presence of bundling interactions [Fig.~\ref{fig:gels}] \cite{pooja2020,martinez2019}. Analogously, contractile systems exhibit splay instabilities, but these have not been experimentally observed. 

The interplay between alignment and tip-accumulation is illustrated at high microtubule concentrations in the absence of bundling interaction [Fig.~\ref{fig:highmtcontract},~\ref{fig:3Dcoral}]. Samples prepared in this regime initially exhibit both aligned filaments and networks contraction. Thus, they are a good candidate for observing the splay instability. Indeed, we observed splay-like deformations, but these were associated with self-tearing. This might be a consequence of the extended nature of microtubule filaments. In polymeric liquid crystals, such as microtubule-based nematics, splay deformations generate local variations in the filament concentration \cite{meyer1984}. Thus, splay instabilities lead to sharp density gradients, which in turn could lead to self-tearing, which yields finite-sized condensates. Beyond this point, the system starts exhibiting structural rearrangements that are likely driven by the tip-accumulation of molecular motors. In particular, the rapidly formed condensates become enveloped by a monolayer of aligned microtubules, which are anchored to a 2D sheet of kinesin motors. The subsequent surface roughening transition is related to the zippering of monolayers into bilayers [Fig.~\ref{fig:3Dcoral}]. It generates dramatic topological rearrangements that transform simple compact condensates into a perforated active foam. Active foams are composed of bilayers, which have both locally aligned filaments and tip accumulated motors. Thus, they resolve the above-described constraints that govern the dynamics of kinesin-4/microtubule systems.

In summary, we demonstrated that kinesin-4 motors self-organize microtubules into a myriad of hierarchical structures. At a single filament level, kinesin-4 motors accumulate at microtubule tips to define a spatially heterogeneous elemental unit capable of higher-order self-assembly. This segmented structure results from a dynamical process, in contrast to amphiphilic systems, where the spatial heterogeneity of the basic building blocks is permanently programmed in the amphiphile’s molecular structure. Tip-decorated microtubules locally condense to generate higher-order radial asters. Asters can, in turn, merge to form extended bilayer sheets. The bilayer sheets form a tissue-like active foam at higher filament concentrations that undergo intriguing motor-driven topological rearrangements. Current hydrodynamic theories do not explain these phenomena. 

\begin{acknowledgments}
We thank Mark Bowick, Boris Shraiman, Linnea Lemma, and Dillon Cislo for valuable discussions. In addition, we thank Shuo Jiang and Marc Ridilla for their assistance in purifying kinesin-4 and Sithara Wijeratne for sharing the results of single-molecule experiments on kinesin-4. DJN acknowledges the support of NSF-DMR-2004380, NSF-DMR-1420570, and NSF-DMR-0820484. RS was supported by a grant from the NIH (1DP2GM126894-01). ZD acknowledges the support of NSF-DMR-2004617 and NSF-MRSEC-2011486. NPM acknowledges support from the Helen Hay Whitney Foundation and NSF PHY-1748958. We also acknowledge the use of Brandeis MRSEC optical microscopy and biosynthesis facilities, which are funded by NSF-MRSEC-2011486.
\end{acknowledgments}

\section{\label{sec:Methods} Methods}

\subsection{\label{sec:SamplePrep} Sample Preparation}
We studied kinesin-4 driven dynamics by combining GFP-labeled kinesin with Alexa-647 labeled stabilized microtubules in a buffered solution with an ATP regeneration system. The solution consisted of DI water with 80 mM PIPES (piperazine-N, N'-bis), 5 mM magnesium chloride, 1 mM EGTA, 1.4 mM ATP (Adenosine triphosphate, Sigma A2383), 0.034\% pyruvate kinase (PK/LDH, Sigma P-0294), and 52 mM PEP (Phosphoenolpyruvate, Alfa Aesar B20358) adjusted to a pH of 6.8 with potassium hydroxide. In addition, to prevent photobleaching, we added DTT (dithiothreitol, ACROS Organics 16568), Glucose (Sigma G7528), Catalase (Sigma C40), and Glucose oxidase (Sigma G2133). When noted, experiments include 35 kDa PEG (polyethylene glycol). 

The full-length human kinesin-4 clone Kif4A or fluorescent Kif4A-GFP were expressed in sf9 cells as described previously \cite{radhika2013}. We purified tubulin from bovine brains according to a previously published protocol \cite{castoldi2003}. This tubulin was polymerized and stabilized into microtubules by mixing 60 uM tubulin with 3 mM of the non-hydrolyzable GTP analog GMPcPP (Guanosine-5’-[($\alpha$,$\beta$)-methyleno]triphosphate, Jena Biosciences NU-405), and a solution of 1 mM DTT, 80 mM PIPES, 2 mM magnesium chloride, 1 mM EGTA in DI water adjusted to a pH of 6.8 with potassium hydroxide. 3\% of tubulin monomers were labeled with a fluorescent dye, Alexa-Fluor 647 (Invitrogen, A-20006), by a succinimidyl ester linker according to a previously published protocol \cite{hyman1991}. The solution was incubated in a water bath at 310 K for one hour and then left to cool to room temperature for 6 hours. Polymerized microtubules were flash-frozen in liquid and subsequently thawed before creating a sample. 

While all active materials consist of GMPcPP polymerized microtubules, the paper’s concentrations refer to tubulin concentrations. A microtubule consists of a repeating lattice of $\sim$13 tubulin monomers, each ring of the lattice is 4 nm \cite{howard2002book}. Thus if the mean microtubule length is approximately 4.9 $\mu$m, each microtubule has roughly 16,000 tubulin monomers. 

\subsection{\label{sec:ChamberPrep} Chamber Preparation}
Each experiment occurs in a chamber with dimensions of 1.5 mm x 0.1 mm x 18 mm unless noted otherwise. The chamber consists of a glass top and bottom, with parafilm spacers sealed with NOA 81 UV Adhesive (Norland Products, 8101) at both ends. The glass was coated with a polyacrylamide brush to suppress proteins’ adsorption onto the glass \cite{lau2009}. To bond parafilm to the glass, we warm the parafilm to 338 K and press it onto the glass with the rounded end of a PRC tube. This process leads to chambers that are 80-100 $\mu$m in height.

\subsection{\label{sec:MTLength} Microtubule Length Distribution Measurements}
To measure microtubule length distributions, we flow dilute microtubules into an untreated glass chamber. Microtubules adsorbed onto the glass are imaged with a 100x objective with a 1.2 NA (Numerical Aperture) and an automated stage. The resulting data set is segmented based on a simple threshold. Each segmented object is then fit to an ellipse. If the ellipse has a thin minor axis compared to its principal axis, then it is recorded as a microtubule with the principal axis as the length. This process discards overlapping or out-of-focus microtubules.

\subsection{\label{sec:Microscopy} Microscopy}
Fluorescence images were captured using a Nikon Ti2 base attached to an Andor Zyla using a 4x Nikon Plan Apo Lambda (NA, 0.2) objective or a 10x Nikon Plan Fluor objective (NA, 0.3).

Confocal microscopy images were captured with a Crest X-Light V2 spinning disk system attached to a Nikon Ti2 base and a Hamamatsu ORCA-Flash4.0 V3. The objective used for the aster sedimentation data was a 40x Plan Fluor objective (NA, 0.75). The objective used for all other data was a 40x Apo long working distance water immersion objective (NA, 1.15). Zeiss Immersol W, an NA matched oil substitute, prevented imaging deterioration due to water evaporation during long acquisitions.

\bibliography{ms}

\clearpage

\onecolumngrid
\begin{center}
\textbf{\large Active microphase separation\\ in mixtures of microtubules and tip-accumulating molecular motors\\Supplementary Material}\\[.2cm]
\end{center}

\setcounter{equation}{0}
\setcounter{figure}{0}
\setcounter{table}{0}
\setcounter{page}{1}
\setcounter{section}{0}
\renewcommand{\theequation}{S\arabic{equation}}
\renewcommand{\thefigure}{S\arabic{figure}}

\section{\label{sec:SuppInfo} Supplementary Information}

\subsection{\label{sec:SuppAsterConvolution} Prediction of Aster and Bilayer Structure}

We analytically match the microtubule intensity profile in asters and bilayers using only the intensity profile of molecular motors and a measured length distribution of microtubules [Fig.~\ref{fig:aster}(c)]. To do this, we model the microtubule profile as arising from microtubules of various lengths attached to the kinesin by their ends. This is equivalent to the convolution of the molecular motor profile with a probability distribution of microtubule length. We represent the measured length distribution of stabilized microtubules with a log-normal distribution $f(\Lambda)$:
\begin{equation} f(\Lambda) = \frac{1}{\Lambda S \sqrt{2 \pi} } \exp \left ( - \frac{(\ln(\Lambda) - M)^2}{2S^2} \right ) , \end{equation} 
where $\Lambda$ is the non-dimensionalized length $L/L_0$, and $M$ and $S$ are fit parameters related to the dimensionless mean $\mu/L_0$ and the variance $\sigma^2/L_0^2$ as:
\begin{equation} \mu/L_0=e^{M+\frac{S^2}{2}} ,\end{equation} 
\begin{equation} \sigma^2/L_0^2=e^{S^2+2M} \left ( e^{S^2} - 1 \right ) .\end{equation} 
The probability that a microtubule has a dimensionless length greater than $d/L_0$ is the integral of that distribution from the length $d/L_0$ out to infinity:
\begin{equation} I_{1d}(d)=\nu_{1d} \int_{d/L_0}^{\infty}f(\Lambda)d\Lambda=\frac{\nu_{1d}}{2}+\frac{\nu_{1d}}{2} erf \left ( \frac{M-\ln(d/L_0)}{\sqrt{2} S} \right ) ,\end{equation} 
where $\nu_{1d}$ is a normalization factor that includes the conversion to fluorescent intensity. This equation represents the normalized microtubule intensity profile of microtubules perpendicularly anchored on one side of a plane. The only fit parameter is the normalization factor $\nu_{1d}$, as all other variables have been extracted from the measured length distribution.

In order to predict the structure of a radially symmetric aster, it is helpful to extend this analysis to radial coordinates. A radially oriented microtubule at a distance $r$ from its kinesin anchor takes on the same form, but with a factor $1/r$: 
\begin{equation} I_{r}(d)=\frac{\nu_{r}}{2\pi r} \int_{r/L_0}^{\infty}f(\Lambda)d\Lambda=\frac{\nu_{r}}{4 \pi r}+\frac{\nu_{r}}{4\pi r} erf \left ( \frac{M-\ln(r/L_0)}{\sqrt{2} S} \right ) ,\end{equation}
where $\nu_{r}$ is a normalization factor adjusted for radial coordinates. Finally, we convolve this result with the imaging point spread function $f_{ps}$, measured from 50 nm fluorescent colloids. 
\begin{equation} I_{MT}^{aster} = I_K^{aster} * I_{r} * f_{ps} \end{equation}
Convolving the distribution $I_r(d)$ with the radial profile of kinesin intensity $I_K^{aster}$ and the point spread function $f_{ps}$ creates the radial aster microtubule intensity profile $I_{MT}^{aster}$ which closely matches the experimental microtubule intensity profile as shown in Figure~\ref{fig:aster}(e).

The equivalent calculation for the contracted bilayer’s microtubule profile, shown in Figure~\ref{fig:globalcontract}(f), can be reduced to a one-dimensional problem. We construct the bilayer microtubule profile as microtubules perpendicularly anchored on a plane and thus use the 1D model for $I_{1d}$ derived earlier. Convolving $I_{1d}$ with the z-profile of kinesin from the bilayer $I_K^{bilayer}$ in two directions, and then convolving that profile with point spread function $f_{ps}$ creates the bilayer microtubule z-profile $I_{MT}^{bilayer}$.
 
\subsection{\label{sec:SuppAsterSegmentation} Aster Segmentation}
The aster data consists of 2-channel z-stacks with a distance of 0.65 $\mu$m between the imaging planes. The length of a pixel is also 0.65 $\mu$m. To measure the volume and aspect ratio of asters [Fig.~\ref{fig:aster}(f),(g)], we segment the kinesin channel through a simple threshold. This binary data set is refined by a Chan-Vase active contour algorithm operating on the original data set. 
 
\subsection{\label{sec:SuppSedimentationHeight} Sedimentation Height}

The sedimentation height $h_{MT}$ is the height from the base of the chamber at which 1/5 of the total microtubule fluorescence is encompassed. To calculate this height we define the cumulative microtubule density function $D(h)$, an integral of material from the floor of the chamber to height $h$, normalized by the total material in the chamber $\rho_{tot}$
\begin{equation} D(h)=\frac{1}{\rho_{tot}} \int_0^h \rho (y) dy . \end{equation} 
The sedimentation height $h_{MT}$ is then the height at which $D(h_{MT}) = 0.2$. The density $\rho_{MT}$ in Figure~\ref{fig:sediment}(e) is the mean density of all material below the height $h_{MT}$. 

\subsection{\label{sec:SuppOrientation} Orientational Order Parameter and Coherency}
We calculate orientation fields from images by identifying the principal spatial derivatives using a structure tensor \cite{bigun2004, rezakhaniha2012}. A structure tensor $\mathbf{T}$ of two-dimensional gradients is constructed from a 3D signal intensity field $I$ as 
\begin{equation} 
\mathbf{T} = \begin{bmatrix}
\partial_x \partial_x I_{xyz} & \partial_x \partial_y I_{xyz} \\ 
\partial_y \partial_x I_{xyz} & \partial_y \partial_y I_{xyz}
\end{bmatrix}.
\end{equation} 
The eigenvalue $\lambda_{min}$ of $\mathbf{T}$ associated with the lowest intensity variation represents the vector $\vec{v}_{min}$ along which the intensity gradients are smallest. The direction of $\vec{v}_{min}$ gives the scalar orientation field used to calculate the orientation distribution function.
The coherency $C$ [Fig.~\ref{fig:highmtcontract}(b)] is defined as the difference between the tensor eigenvalues normalized by their sum: 
\begin{equation} C=\frac{\lambda_{max}-\lambda_{min}}{\lambda_{max}+\lambda_{min}}.\end{equation} 
We calculate a field of local orientations $\theta$ from the local values of $\vec{v}_{min}$. The contractions analyzed display negligible bend in their structure, so we define a single average director $\bar{\theta}$ for the entire material as the mean value of $\theta$:
\begin{equation} \bar{\theta} = \frac{1}{N} \sum_{i=1}^N \theta_i .\end{equation} 
From this we calculate the orientational order parameter $S$, defined as:
\begin{equation} S = \langle \cos (2[\theta - \bar{\theta}]) \rangle . \end{equation} 
At late times, microtubule bundles appear anchored normal to the surface. We exclude in the calculation of $\bar{\theta}$ and the orientational order parameter $S$ by using a mask. The mask is generated from a probability field $P_{in}$ using iLastik for pixel classification \cite{ilastik}. 

\subsection{\label{sec:SuppSurfaces} Surface Construction}
To construct numerical surfaces, we start by acquiring confocal data such that each voxel is isotropic. These voxels are classified as “inside” or “outside” the structure of interest by using iLastik to generate a probability field $P_{in}$. Then a binary field $F$ is generated from $P_{in}$ using a morphological snake method. Next a polygonal surface $S$ is constructed from $F$ using a marching cubes algorithm. Finally, the surface $S$ is remeshed at a specified triangle size using Meshlab \cite{meshlab}. Code for this process is available upon request. 
 
\subsection{\label{sec:SuppCorrelation} Normal-normal correlation $C(r)$}
\begin{figure*}
\includegraphics[scale=0.95]{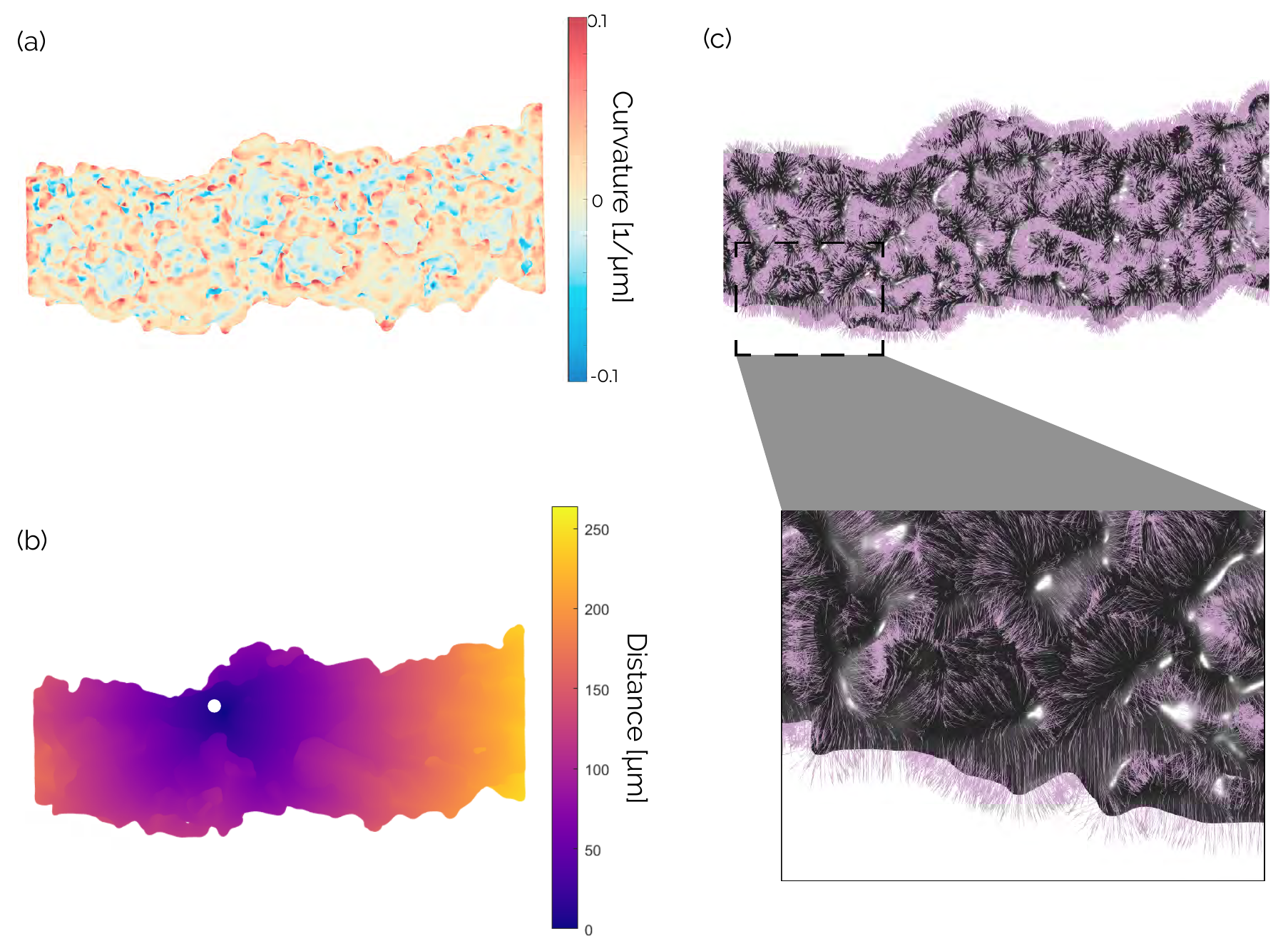}
\caption{\label{fig:SuppNormalNormal}
During roughening, we quantify the normal-normal correlation as a function of geodesic distance along the material surface.
(a) Mean curvature of a globally contracting surface at late time.
(b) Geodesic distance, on the material surface, from an initial point indicated by a white circle.
(c) Lilac arrows indicate normal vectors on the material surface, a random sampling of 10\% of the normal vectors are displayed. The ends of the material along the long axis are cropped off for the calculation of normals.}
\end{figure*}

To determine the normal-normal correlation of a structure we first generate a surface for that structure as described above. We then bisect the surface along the smallest moment of the material. This bisection is to exclude anticorrelations in $C(r)$ due to the curvature of the surface. We calculate a normal vector $\hat{n}(r,t)$ at each point $r$ on the two surface halves at time $t$. The normal-normal correlation is calculated as 
\begin{equation} C(\Lambda,t) 
= \frac{\langle \hat{n}(r,t) \cdot \hat{n}(r+\Lambda,t) \rangle }{\langle \hat{n}(r,t) \cdot \hat{n}(r,t) \rangle } 
= \frac{1}{N_i N_\Lambda} \sum_i^{N_i} \sum_\Lambda^{N_\Lambda} \frac{\hat{n}(r_i,t) \cdot \hat{n}(r_i+\Lambda,t)}{\hat{n}(r_i,t) \cdot \hat{n}(r_i,t)} , \end{equation} 
where angular brackets indicate a spatial average over all initial points $i$ and all geodesic paths $\Lambda$. We calculate geodesics on each half of the surface via a fast-marching mesh algorithm. Figure~\ref{fig:SuppNormalNormal} shows the geodesic distance from a point along a contracted surface and the normal vectors of that contracted surface. Binning by lengths of the path $\Lambda$ at a particular time $t$, we calculate $C(r)$. At small length scales, the normal-normal correlation is reasonably well fit by an exponential. The correlation length [Fig.~\ref{fig:3Dcontractflux}(g), Fig.~\ref{fig:3Dcoral}(d)] is defined as the inverse of the exponent to this fit. Code to generate normals and calculate geodesic distances is available upon request. 

\subsection{\label{sec:SuppContraction} Contraction Kinematics}
If the mass of proteins is conserved, there are constraints relating shape change with protein flux. We consider an enclosed network with a volume $V$ and a boundary of area $A$. The total mass $M$ is the sum of the areal surface density $\rho_A$, plus the sum of the volumetric density $\rho_V$ over the volume:
\begin{equation} M =\int_A \rho_S dA +\int_V \rho_V dV .\end{equation} 
Assuming mass conservation, the time derivative of this quantity is zero. 
That is,
\begin{equation} 0= \partial_t M = \int_A (\partial_t \rho_A) dA + \langle \rho _A \rangle \partial_t A + \int_V( \partial_t \rho_V)dV +\langle \rho_V\rangle \partial_t V ,\end{equation} 
where angular brackets indicate a spatial average. 
Given that protein is found only on the surface and in the bulk, an increase in the first two terms would signal a flux of material from the bulk to the surface, whereas an increase in the second two terms would signal a flux of material into the bulk. 
That is, the net flux of protein from the bulk $V$ to the surface $S$ is
\begin{equation} \Phi_{V\rightarrow S} = A\partial_t \langle \rho_A \rangle + \langle \rho_A \rangle \partial_t A \end{equation} 
while the net flux of protein from the surface to the bulk is
\begin{equation} \Phi_{S\rightarrow V} = V\partial_t \langle \rho_V \rangle + \langle \rho_V \rangle \partial_t V. \end{equation} 

\subsection{\label{sec:SuppPIV} Mean Network Speed and Velocity Correlation Length of K4 Driven Gels}
\begin{figure*}
\includegraphics[scale=1]{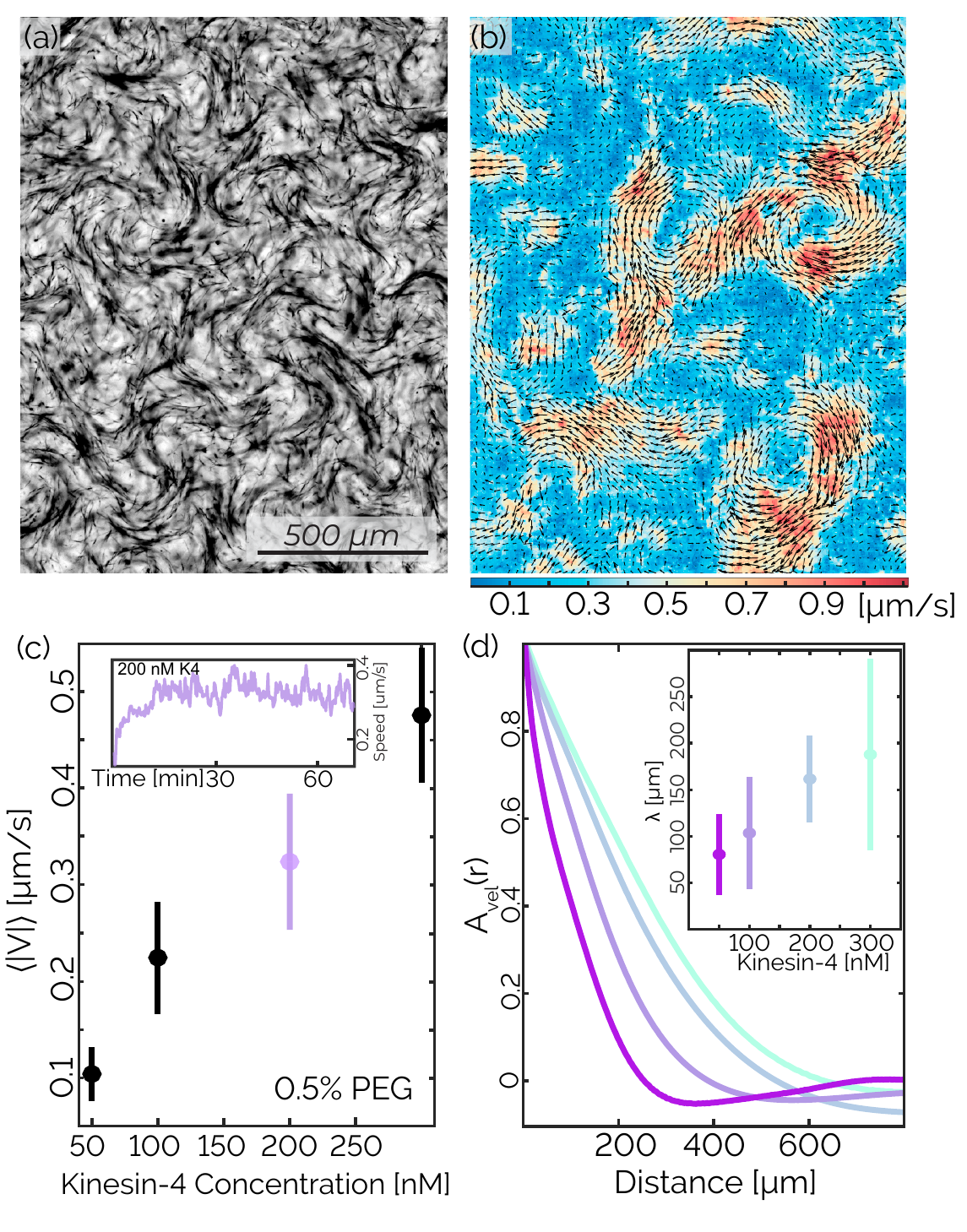}
\caption{\label{fig:SuppPIVGel}
Increasing kinesin concentration amplifies the dynamics of an extensile fluid.
(a) An extensile network driven by kinesin. Sample is created with 200 nM kinesin, 13 $\mu$M tubulin (imaged), 0.5\% PEG.
(b) Color map indicating the magnitude of the material velocity in the previous panel, with overlaid arrows representing the velocity vector field $\vec{v}(r)$.
(c) Mean speed $\langle|V|\rangle$ as a function of kinesin. The error bars are standard deviation (n=3). 
Inset) Spatially averaged speed $U(t)=\langle \vec{v}(r,t)\rangle$ plotted over time for the experiment shown in panel (a).
(d) Time-averaged velocity autocorrelation $A(r)$ as a function of kinesin.
Inset) Length scale $\lambda$ at which $2A(\lambda)=A(0)$. Error bars are standard deviation (n=3).}
\end{figure*}

The velocity field, $v(r,t)$, of the extensile fluid phase was calculated using the velocimetry package PIVLab [Fig.~\ref{fig:SuppPIVGel}(a),(b)] \cite{PIVlab}. From this data we calculated the the mean network speed $\langle \left | V \right | \rangle$ defined as
\begin{equation} \langle \left | V \right | \rangle = \frac{1}{T_f - T_i} \sum_{t=T_i}^{T_f}\langle v(r,t) \rangle \end{equation} 
where $T_f$ is the final time, and $T_i$ indicates time shortly after the initial gel buckling instability. The average inside the sum is over space as defined by the variable $r$. Titrating over kinesin concentration, we found that the mean microtubule network speed $\langle \left | V \right | \rangle$ increased with kinesin concentration [Fig~\ref{fig:SuppPIVGel}(c)].

We used the velocity field $v(r,t)$ to generate a spatial velocity-velocity correlation $A_{vel} (r)$ defined as
\begin{equation}A_{vel}(r)
=\frac{1}{T}\sum_t^T \langle A(r,t) \rangle
= \left \langle \frac{\langle v(r,t) \cdot v(r',t) \rangle }{\langle v(r',t) \cdot v(r',t) \rangle} \right \rangle
= \frac{2}{TN(N-1)} \sum_t^T \sum_i^N \sum_{j<i}^N \frac{v(r_i,t)\cdot v(r_j,t) }{v(r_j,t)\cdot v(r_j,t)}
\end{equation} 
where $T$ is the number of frames evaluated. Here the average inside the sum is over space as defined by the variable $r'$. This correlation was evaluated in Fourier space to reduce computation time. We measured a correlation length scale $\lambda$, defined as the length scale at which $A_{vel}(r)$ has decayed to half of its initial amplitude. In contrast to studies of truncated kinesin-1, increasing kinesin-4 concentration increased the velocity-velocity correlation length scale $\lambda$ [Fig.~\ref{fig:SuppPIVGel}(d)] \cite{gil2014}.

\subsection{\label{sec:SuppBoundaries} Modifications of contraction phenomena due to boundary conditions}
\begin{figure*}
\includegraphics[width=\textwidth]{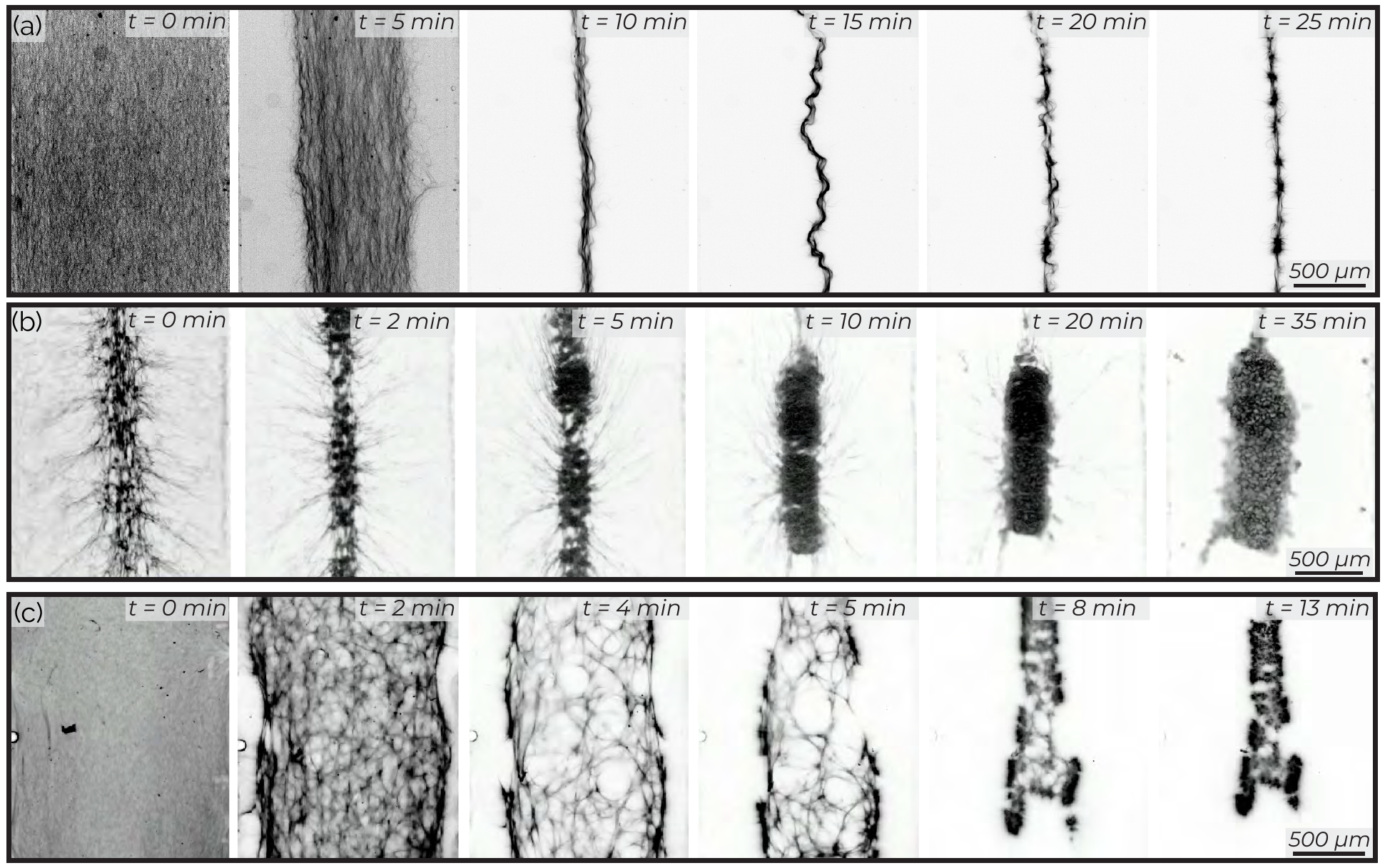}
\caption{\label{fig:SuppBoundaryConditions}
The behavior of a globally contracting system is influenced by the conditions at the borders of the chamber.
(a) A contracting material pinned at the ends of the chamber many millimeters away. This material first contracts but then buckles at 15 min, followed by straightening again at 20 min.
(b) A global contraction with some sticking at the parafilm chamber edges. This sample forms an active bilayer foam as its end state.
(c) A global contraction loses the symmetry imparted on it by the chamber.}
\end{figure*}

The dynamics and final structure of a global contraction are sensitive to microtubule concentration, kinesin concentration, initial microtubule alignment, and boundary conditions. When the material is pinned at the ends of the chamber the resulting global contraction displays significant phenomenological differences from its unpinned form [Fig.~\ref{fig:SuppBoundaryConditions}(a)]. First contracting material pinned at the ends of the chamber contract to a thin line. Then the line of material buckles, and then the line of material straightens again. Finally, at long time scales, material accumulates at intervals along the line of the contraction, forming large aster-like clumps. Similarly, non-specific adhering to the chamber sides changed the form of the contraction [Fig.~\ref{fig:SuppBoundaryConditions}(b), (c)].

\section{\label{sec:SuppVideos} Supplementary Videos}

\begin{itemize}

\item Video 1: At low microtubule concentrations, dynamic asters spontaneously assemble. This video shows orthogonal planes projected over 6.5 $\mu$m. Sample is created with 200 nM kinesin-4 (blue), 400 nM tubulin (black).

\item Video 2: At intermediate microtubule density, networks of microtubules globally contract. This video shows four fields of epifluorescent imaging of fluorescent microtubules stitched together. The sample is created with 50 nM kinesin-4 (blue), 1000 nM tubulin (black).

\item Video 3: Sedimented asters do not merge and globally contract. This video shows a 3D projection from confocal stacks, with two xy slices at indicated positions. These slices are z-projected over 6.5 $\mu$m. Sample is created in a 300 $\mu$m chamber with 200 nM kinesin-4 (blue), 400 nM tubulin (black). 

\item Video 4: At high microtubule density, global contractions align and then roughen. This video shows a z-projection from confocal data, along with xz and yz orthogonal slices projected over 6.5 $\mu$m. Starting at 110 min, a 3D surface (blue) generated from the dense surface of the condensate is displayed. This surface shows the location of the orthogonal slices. Sample consisted of 10 $\mu$M tubulin (black) and 200 nM kinesin (blue).

\item Video 5: At the highest microtubule density, kinesin-4 drives microtubule condensation and the subsequent formation of an active foam. This video shows a 3D projection from confocal stacks of a 333x333x100 $\mu$m field of view. Intermittent pauses in the video show the interior structure of the material during its development. Sample contained 200 nM kinesin (blue), 40 $\mu$M tubulin (black).

\item Video 6: Whole-chamber epifluorescent imaging of highest-density microtubule systems buckling, condensing, and forming an active foam. Sample constituted from 200 nM kinesin (blue), 40 $\mu$M tubulin (black).

\item Video 7: A series of videos shows a titration of microtubule concentrations in the presence of PEG, resulting in a transition from extensile to contracting networks. All videos are epifluorescent imaging of fluorescent microtubules.

\item Video 8: Two videos of extensile networks transforming into bilayer structures. The first video is epifluorescent imaging of fluorescent microtubules (black) and kinesin (blue). The second video is a max-z projection of confocal imaging of a sample in a thin (30 $\mu$m) chamber.

\end{itemize}

\section{\label{sec:SuppFigures} Supplementary Figures}

\begin{figure*}
\includegraphics[scale=1.5]{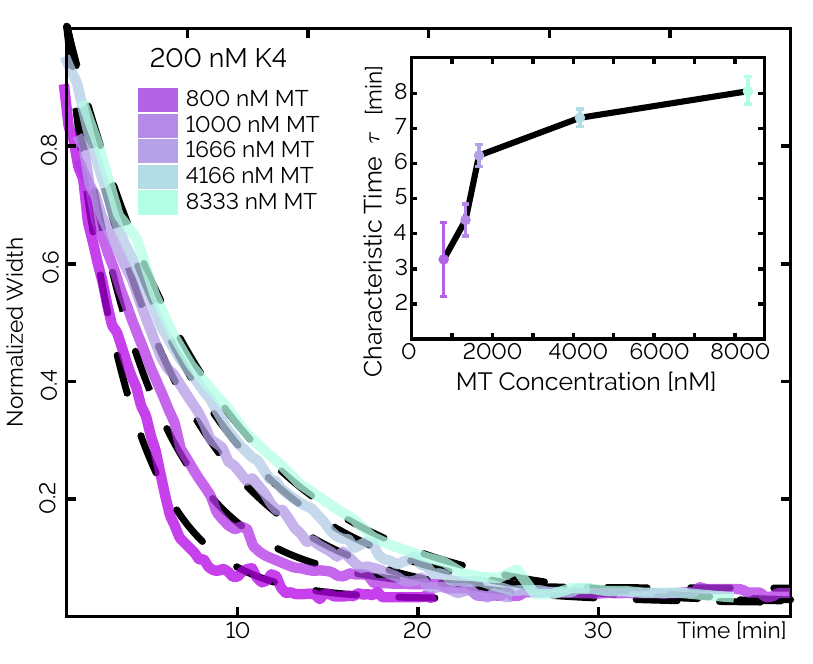}
\caption{\label{fig:SuppContract}
The contraction time scale decreased with increasing microtubule number density.
Plotted is the normalized width $W_n$ at five tubulin concentrations (200 nM kinesin). Dashed lines represent the exponential fit $f_c (t)$.
Inset) Characteristic time $\tau$ for each tubulin concentration.}
\end{figure*}

\begin{figure*}
\includegraphics[scale=1]{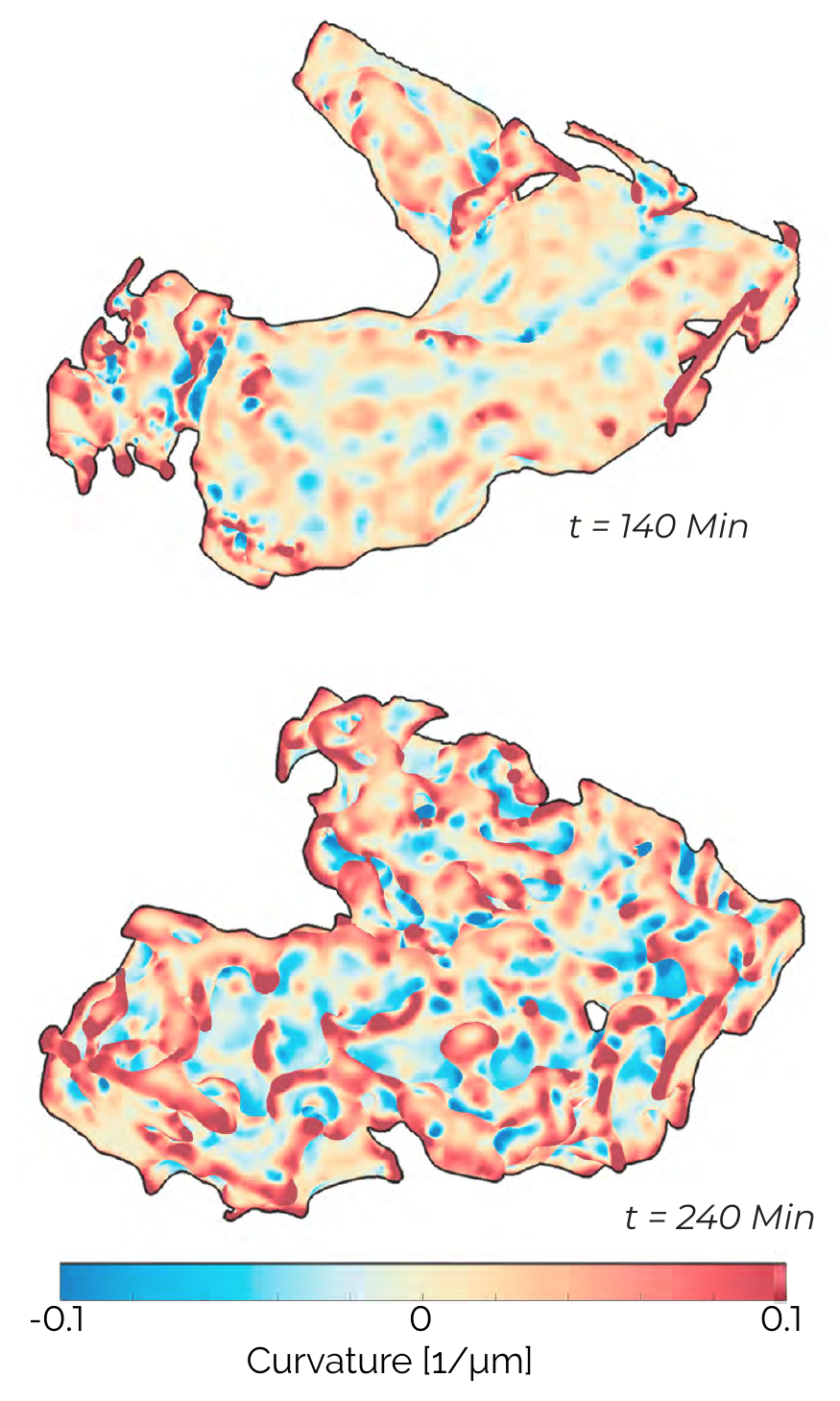}
\caption{\label{fig:SuppMean} 
The mean curvature of the condensate surface shown in Fig.~\ref{fig:3Dcoral} at two points in time, the first after early monolayer formation, the second at the onset of bilayer formation.}
\end{figure*}

\begin{figure*}
\includegraphics[scale=1]{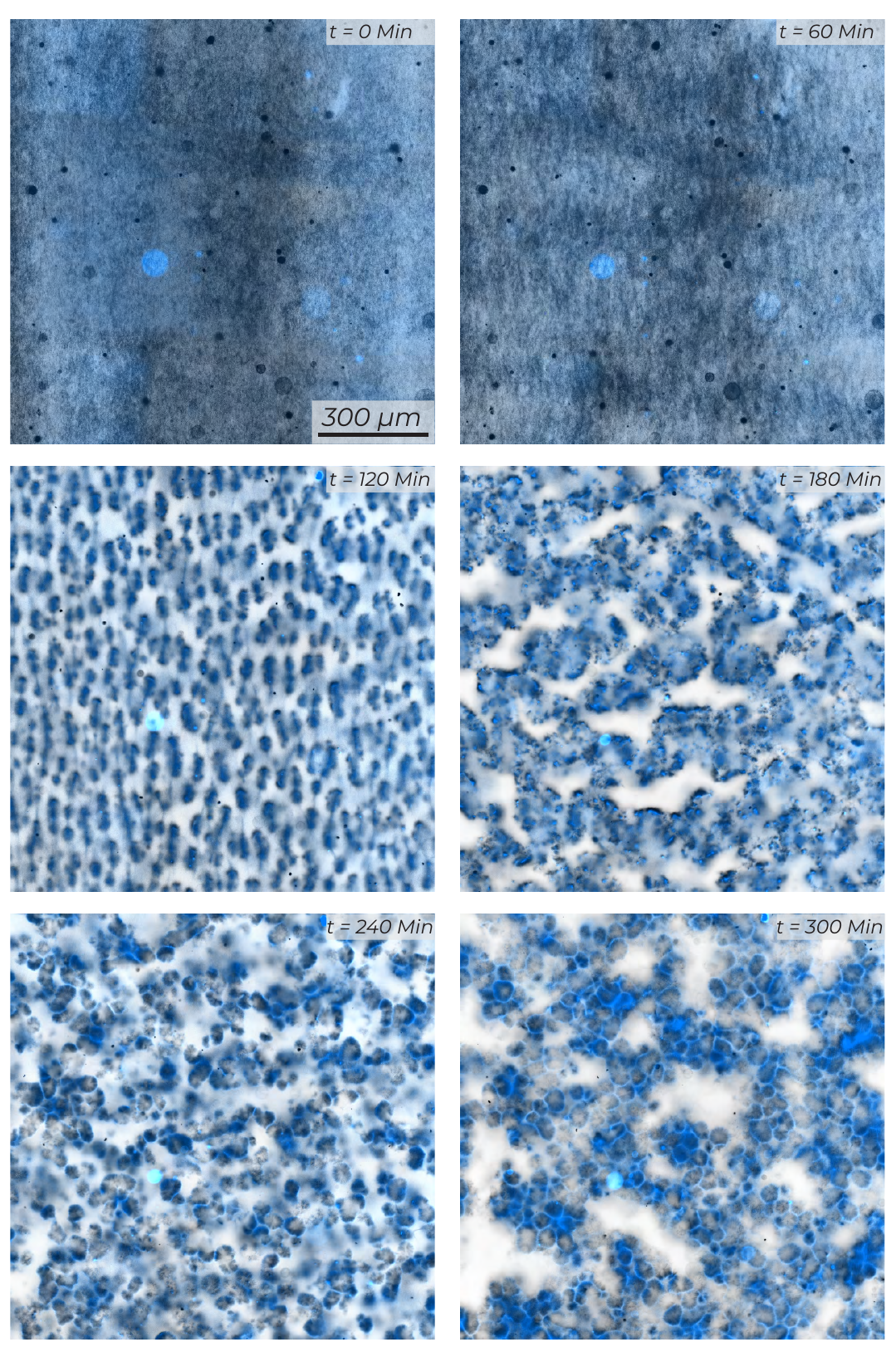}
\caption{\label{fig:SuppFoamEvolution}
Low magnification imaging shows the slight buckling and splay of microtubules into monolayer envelopes, followed by the deformation of monolayers into an active bilayer foam.}
\end{figure*}

\begin{figure*}
\includegraphics[width=\textwidth]{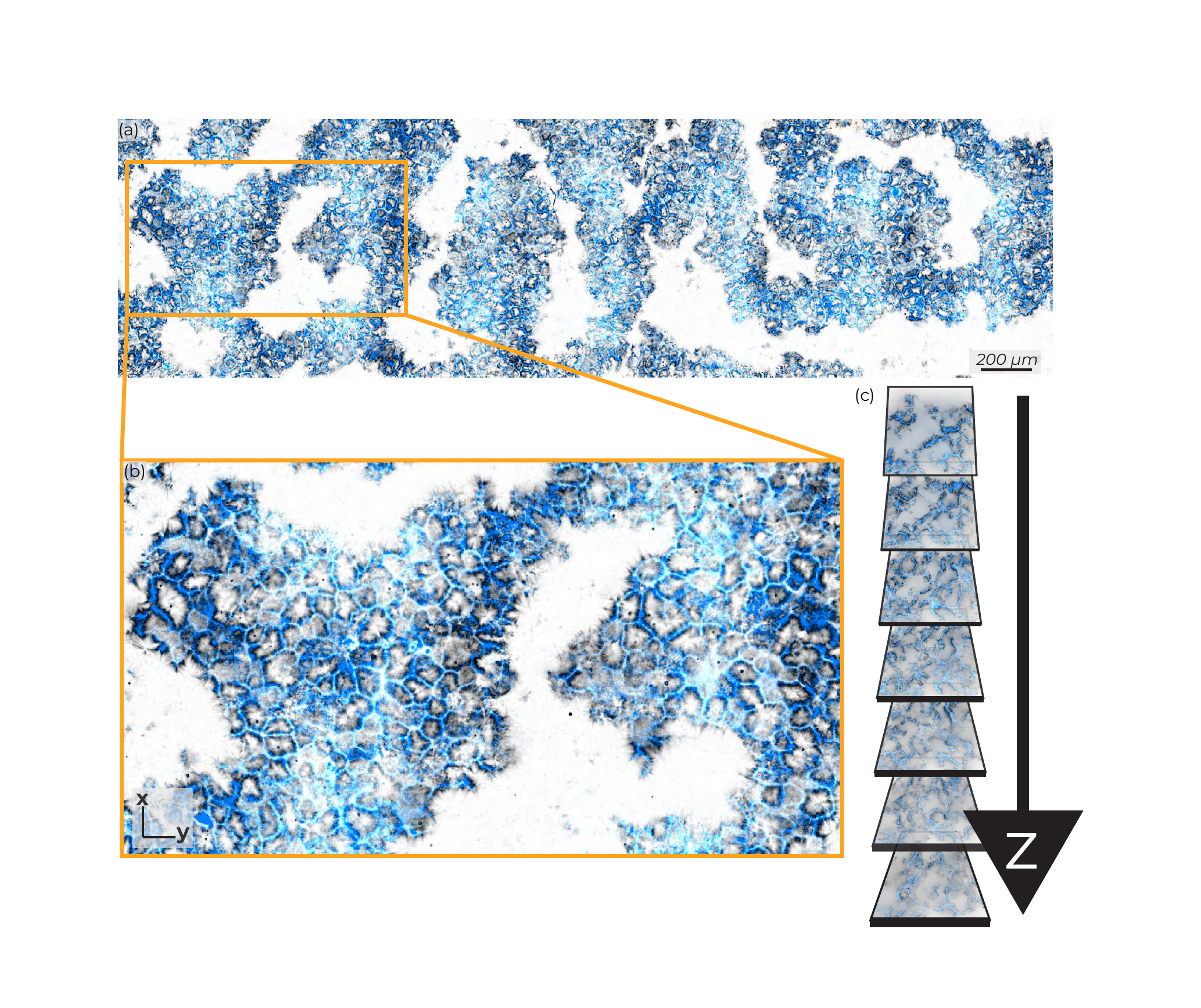}
\caption{\label{fig:SuppBigFoam}
At high MT concentrations, the mixture coarsens into an active foam.
(a) Maximum intensity projection over 10 $\mu$m in z of an entire chamber of foam.
(b) zoom in (c) Z-stack of 6.5 $\mu$m z-projection slices with an additional 6.5 $\mu$m in between each slice, showing the 3D structure of a bilayer foam.}
\end{figure*}

\end{document}